\documentclass[twocolumn]{aastex62}
\usepackage[utf8]{inputenc}
\submitjournal{ApJ}
\accepted{March 29, 2021}

\begin{document}


\title{The JCMT BISTRO Survey: The Distribution of Magnetic Field Strengths towards the OMC-1 Region}

\author[0000-0001-7866-2686]{Jihye Hwang}
\email{hjh3772@gmail.com}
\affil{Korea Astronomy and Space Science Institute (KASI), 776 Daedeokdae-ro, Yuseong-gu, Daejeon 34055, Republic of Korea}
\affil{University of Science and Technology, Korea (UST), 217 Gajeong-ro, Yuseong-gu, Daejeon 34113, Republic of Korea}

\author[0000-0002-1229-0426]{Jongsoo Kim}
\affil{Korea Astronomy and Space Science Institute (KASI), 776 Daedeokdae-ro, Yuseong-gu, Daejeon 34055, Republic of Korea}
\affil{University of Science and Technology, Korea (UST), 217 Gajeong-ro, Yuseong-gu, Daejeon 34113, Republic of Korea}

\author[0000-0002-8557-3582]{Kate Pattle}
\affil{Centre for Astronomy, National University of Ireland Galway, University Road, Galway, Ireland}

\author[0000-0003-4022-4132]{Woojin Kwon}
\affil{Department of Earth Science Education, Seoul National University (SNU), 1 Gwanak-ro, Gwanak-gu, Seoul 08826, Republic of Korea}
\affil{Korea Astronomy and Space Science Institute (KASI), 776 Daedeokdae-ro, Yuseong-gu, Daejeon 34055, Republic of Korea}

\author[0000-0001-7474-6874]{Sarah Sadavoy}
\affil{Department for Physics, Engineering Physics and Astrophysics, Queen's University, Kingston, ON, K7L 3N6, Canada}

\author[0000-0003-2777-5861]{Patrick M. Koch}
\affil{Academia Sinica Institute of Astronomy and Astrophysics, No. 1, Sec. 4., Roosevelt Road, Taipei 10617, Taiwan}

\author[0000-0002-8975-7573]{Charles L. H. Hull}
\affil{National Astronomical Observatory of Japan, NAOJ Chile, Alonso de C\'ordova 3788, Office 61B, 7630422, Vitacura, Santiago, Chile }
\affil{Joint ALMA Observatory, Alonso de C\'ordova 3107, Vitacura, Santiago, Chile}
\affil{NAOJ Fellow}

\author[0000-0002-6773-459X]{Doug Johnstone}
\affil{NRC Herzberg Astronomy and Astrophysics, 5071 West Saanich Road, Victoria, BC V9E 2E7, Canada}
\affil{Department of Physics and Astronomy, University of Victoria, Victoria, BC V8W 2Y2, Canada}

\author[0000-0003-0646-8782]{Ray S. Furuya}
\affil{Institute of Liberal Arts and Sciences Tokushima University, Minami Jousanajima-machi 1-1, Tokushima 770-8502, Japan}

\author[0000-0002-3179-6334]{Chang Won Lee}
\affil{Korea Astronomy and Space Science Institute (KASI), 776 Daedeokdae-ro, Yuseong-gu, Daejeon 34055, Republic of Korea}
\affil{University of Science and Technology, Korea (UST), 217 Gajeong-ro, Yuseong-gu, Daejeon 34113, Republic of Korea}

\author[0000-0002-1959-7201]{Doris Arzoumanian}
\affil{Instituto de Astrof\'isica e Ci{\^e}ncias do Espa\c{c}o, Universidade do Porto, CAUP, Rua das Estrelas, PT4150-762 Porto, Portugal}

\author[0000-0001-8749-1436]{Mehrnoosh Tahani}
\affil{Dominion Radio Astrophysical Observatory, Herzberg Astronomy and Astrophysics Research Centre, National Research Council Canada, P. O. Box 248, Penticton, BC V2A 6J9 Canada}

\author[0000-0003-4761-6139]{Chakali Eswaraiah}
\affil{CAS Key Laboratory of FAST, National Astronomical Observatories, Chinese Academy of Sciences, People's Republic of China; University of Chinese Academy of Sciences, Beijing 100049, People's Republic of China}
\affil{National Astronomical Observatories, Chinese Academy of Sciences, A20 Datun Road, Chaoyang District, Beijing 100012, People's Republic of China}

\author[0000-0002-5286-2564]{Tie Liu}
\affil{Key Laboratory for Research in Galaxies and Cosmology, Shanghai Astronomical Observatory, Chinese Academy of Sciences, 80 Nandan Road, Shanghai 200030, People's Republic of China}

\author[0000-0002-3036-0184]{Florian Kirchschlager}
\affil{Department of Physics and Astronomy, University College London, WC1E 6BT London, UK}

\author[0000-0003-2412-7092]{Kee-Tae Kim}
\affil{Korea Astronomy and Space Science Institute (KASI), 776 Daedeokdae-ro, Yuseong-gu, Daejeon 34055, Republic of Korea}

\author[0000-0002-6510-0681]{Motohide Tamura}
\affil{Department of Astronomy, Graduate School of Science, The University of Tokyo, 7-3-1 Hongo, Bunkyo-ku, Tokyo 113-0033, Japan}
\affil{Astrobiology Center, 2-21-1 Osawa, Mitaka-shi, Tokyo 181-8588, Japan}
\affil{National Astronomical Observatory, 2-21-1 Osawa, Mitaka-shi, Tokyo 181-8588, Japan}

\author[0000-0003-2815-7774]{Jungmi Kwon}
\affil{Department of Astronomy, Graduate School of Science, The University of Tokyo, 7-3-1 Hongo, Bunkyo-ku, Tokyo 113-0033, Japan}

\author[0000-0002-9907-8427]{A-Ran Lyo}
\affil{Korea Astronomy and Space Science Institute (KASI), 776 Daedeokdae-ro, Yuseong-gu, Daejeon 34055, Republic of Korea}

\author[0000-0002-6386-2906]{Archana Soam}
\affil{SOFIA Science Center, Universities Space Research Association, NASA Ames Research Center, Moffett Field, California 94035, USA}

\author[0000-0001-7379-6263]{Ji-hyun Kang}
\affil{Korea Astronomy and Space Science Institute (KASI), 776 Daedeokdae-ro, Yuseong-gu, Daejeon 34055, Republic of Korea}

\author[0000-0001-7491-0048]{Tyler L. Bourke}
\affil{SKA Organisation, Jodrell Bank, Lower Withington, Macclesfield, SK11 9FT, UK}

\author[0000-0002-6906-0103]{Masafumi Matsumura}
\affil{Faculty of Education \& Center for Educational Development and Support, Kagawa University, Saiwai-cho 1-1, Takamatsu, Kagawa, 760-8522, Japan}

\author[0000-0002-6956-0730]{Steve Mairs}
\affil{East Asian Observatory, 660 N. A'oh\={o}k\={u} Place, University Park, Hilo, HI 96720, USA}

\author[0000-0003-2011-8172]{Gwanjeong Kim}
\affil{Nobeyama Radio Observatory, National Astronomical Observatory of Japan, National Institutes of Natural Sciences, Nobeyama, Minamimaki, Minamisaku, Nagano 384-1305, Japan}

\author[0000-0001-8467-3736]{Geumsook park}
\affil{Korea Astronomy and Space Science Institute (KASI), 776 Daedeokdae-ro, Yuseong-gu, Daejeon 34055, Republic of Korea}

\author[0000-0001-5431-2294]{Fumitaka Nakamura}
\affil{Division of Theoretical Astronomy, National Astronomical Observatory of Japan, Mitaka, Tokyo 181-8588, Japan}
\affil{SOKENDAI (The Graduate University for Advanced Studies), Hayama, Kanagawa 240-0193, Japan}

\author[0000-0002-8234-6747]{Takashi Onaka}
\affil{Department of Physics, Faculty of Science and Engineering, Meisei University, 2-1-1 Hodokubo, Hino, Tokyo 1191-8506, Japan}
\affil{Department of Astronomy, Graduate School of Science, The University of Tokyo, 7-3-1 Hongo, Bunkyo-ku, Tokyo 113-0033, Japan}

\author[0000-0002-4154-4309]{Xindi Tang}
\affil{Xinjiang Astronomical Observatory, Chinese Academy of Sciences, 830011 Urumqi, People's Republic of China}

\author[0000-0003-3343-9645]{Hong-Li Liu}
\affil{Chinese Academy of Sciences, South America Center for Astrophysics, Camino El Observatorio \#1515, Las Condes, Santiago, Chile}
\affil{Shanghai Astronomical Observatory, Chinese Academy of Sciences, 80 Nandan Road, Shanghai 200030, People's Republic of China}

\author[0000-0003-1140-2761]{Derek Ward-Thompson}
\affil{Jeremiah Horrocks Institute, University of Central Lancashire, Preston PR1 2HE, United Kingdom}

\author[0000-0003-3010-7661]{Di Li}
\affil{NAOC-UKZN Computational Astrophysics Centre, University of KwaZulu-Natal, Durban 4000, South Africa}

\author[0000-0003-2017-0982]{Thiem Hoang}
\affil{Korea Astronomy and Space Science Institute (KASI), 776 Daedeokdae-ro, Yuseong-gu, Daejeon 34055, Republic of Korea}
\affil{University of Science and Technology, Korea (UST), 217 Gajeong-ro, Yuseong-gu, Daejeon 34113, Republic of Korea}

\author[0000-0003-1853-0184]{Tetsuo Hasegawa}
\affil{National Astronomical Observatory of Japan, National Institutes of Natural Sciences, Osawa, Mitaka, Tokyo 181-8588, Japan}

\author[0000-0002-5093-5088]{Keping Qiu}
\affil{School of Astronomy and Space Science, Nanjing University, 163 Xianlin Avenue, Nanjing 210023, Peopl$e'$s Republic of China}
\affil{Key Laboratory of Modern Astronomy and Astrophysics (Nanjing University), Ministry of Education, Nanjing 210023, Peopl$e'$s Republic of China}

\author[0000-0001-5522-486X]{Shih-Ping Lai}
\affil{Institute for Astronomy and Department of Physics, National Tsing Hua University, No. 101, Sec. 2, Guangfu Road, Hsinchu 30013, Taiwan}
\affil{Academia Sinica Institute of Astronomy and Astrophysics, No. 1, Sec. 4., Roosevelt Road, Taipei 10617, Taiwan}

\author[0000-0002-0794-3859]{Pierre Bastien}
\affil{Centre de recherche en astrophysique du Qu$\acute{e}$bec \& d$\acute{e}$partement de physique, Universit$\acute{e}$ de Montr$\acute{e}$al, C.P. 6128 Succ. Centre-ville, Montr$\acute{e}$al, QC, H3C 3J7, Canada}

\begin{abstract}

Measurement of magnetic field strengths in a molecular cloud is essential for determining the criticality of magnetic support against gravitational collapse. In this paper, as part of the JCMT BISTRO survey, we suggest a new application of the Davis-Chandrasekhar-Fermi (DCF) method to estimate the distribution of magnetic field strengths in the OMC-1 region.
We use observations of dust polarization emission at 450 $\mu$m and 850 $\mu$m, and C$^{18}$O (3-2) spectral line data obtained with the JCMT. We estimate the volume density, the velocity dispersion and the polarization angle dispersion in a box, 40$\arcsec\times$40$\arcsec$ (5$\times$5 pixels), which moves over the OMC-1 region. By substituting three quantities in each box to the DCF method, we get magnetic field strengths over the OMC-1 region. We note that there are very large uncertainties in inferred field strengths, as discussed in detail in this paper. The field strengths vary from 0.8 to 26.4 mG and their mean value is about 6 mG.
Additionally, we obtain maps of the mass-to-flux ratio in units of a critical value and the Alfv$\acute{e}$n mach number. The central parts of the BN-KL and South (S) clumps in the OMC-1 region are magnetically supercritical, so the magnetic field cannot support the clumps against gravitational collapse. However, the outer parts of the region are magnetically subcritical. The mean Alfv$\acute{e}$n mach number is about 0.4 over the region, which implies that the magnetic pressure exceeds the turbulent pressure in the OMC 1 region.

\end{abstract}


\section{Introduction} \label{sec:intro}

Magnetic fields and turbulence in star-forming regions have been considered as mechanisms for supporting molecular clouds against gravitational collapse to explain the low star formation efficiency. It is observationally well-known that the lifetimes of molecular clouds are typically longer than their free-fall timescales (\citealt{Hartmann2001}; \citealt{Padoan2014}; \citealt{Federrath2016}) and that star formation efficiency in molecular clouds is observed to be about 1 - 9\% (e.g., \citealt{Carpenter2000}; \citealt{Goldsmith2008}; \citealt{Evans2009}; \citealt{Dunham2015}). In the central part of dense core/clump whose size is less than 0.1 pc, due to the low-ionization fraction, there prevails an efficient decoupling between neutral and charged particles. Because of this, neutral particles move inward the cloud and drag charged particles as well as magnetic field lines \citep{Mouschovias2006}. 
 This is ambipolar diffusion. The resulting magnetic field distribution shows an ordered field geometry such as an hourglass morphology \citep{Galli1993}. The ambipolar diffusion timescale in a magnetized dense core/clump is one or two orders longer than the free-fall timescale (e.g., \citealt{Mestel1956}; \citealt{Nakano1972}; \citealt{LiNaka2004}; \citealt{Nakamura2008}), which could explain the low star formation efficiency. Turbulence also resists the gravitational collapse in molecular clouds. The preferred locations of star formation in molecular clouds are places where convergent flows driven by large-scale turbulence meet (e.g., \citealt{MacLow2004}). The lifetimes of molecular clouds supported by turbulence are comparable to or slightly longer than their free-fall timescales (e.g., \citealt{Vazquez2005}). Although there have been many theoretical studies aimed at understanding the support mechanism and the low star formation efficiency of a molecular cloud, the relative importance of magnetic fields with respect to turbulence is still under debate.

In magnetized molecular clouds, elongated dust grains are aligned with respect to magnetic field lines (e.g., \citealt{Andersson2015}).
The current leading theory of dust grain alignment, radiative alignment torques (RATs), suggests that a spinning dust grain in a molecular cloud is aligned with its minor axis parallel to the magnetic field direction \citep{Lazarian2007}. When dust grains are aligned by RATs, the dust grains absorb the background starlight at optical and near-infrared wavelengths. The direction of polarization at these wavelengths is along the magnetic field direction due to the relatively higher extinction by dust grains along their major axes. 
Conversely, the direction of the dust polarization from the thermal dust emission at far-infrared and sub-millimeter wavelengths will be along the major axes of the dust grains, and so perpendicular to the magnetic field direction. 

Polarization observations are commonly used to trace magnetic field structures and show that the magnetic fields play an important role in star-forming regions. For example, polarized emission at near-infrared wavelengths is well-ordered and perpendicular to the Heiles Cloud 2 and the B211/B213 filaments in the Taurus molecular cloud (\citealt{Tamura1987}; \citealt{Chapman2011}; \citealt{Palmeirim2013}). Using the histogram of relative orientations technique, \citet{Soler2013} showed that the direction of the magnetic field formed in the highly magnetized cloud in their numerical simulations is perpendicular to the dense filaments, while low density filaments are parallel to the magnetic field lines. Some studies have shown an ordered hourglass morphology which is predicted by strong magnetic field models in entire clouds (e.g., \citealt{Sugitani2011}), in single-dish observations (e.g., \citealt{Matthews2009}; \citealt{Ward_Thompson2017}; \citealt{Chuss2019}), and at interferometer/core-envelope scales (e.g., \citealt{Coppin2000}; \citealt{Girart2006}; \citealt{Rao2009}; \citealt{Tang2009}; \citealt{Stephens2013}; \citealt{Hull2014}; \citealt{Kwon2019}; \citealt{Hull2020}). The magnetic field structure in the spheroidal model of a star-forming cloud shows an hourglass shape by increasing density ratio of the cloud (e.g., \citealt{Myers2018}; \citealt{Myers2020}). 
The ordered structure of the magnetic field suggests that the magnetic field plays an important role in a star-forming region. However, the structure alone is not sufficient to judge whether or not the magnetic field can resist the gravitational collapse, so it is crucial to measure magnetic field strengths in a star-forming region.

Magnetic field strengths can be determined using the Davis-Chandrasekhar-Fermi (DCF) method (\citealt{Davis1951}; \citealt{ChanFer1953}). In the original DCF method, the strength of a uniform magnetic field perturbed by non-thermal motions could be estimated by measuring the polarization angle dispersion, the velocity dispersion and the number density of a star-forming region (e.g., \citealt{Crutcher2004}). 
\citet{Ostriker2001} obtained polarization maps along different lines of sight from numerical simulations and modified the original DCF formula by a factor 0.5. They restricted the validity of the method to cases where the polarization angle dispersion is less than 25$^{\circ}$. The DCF method suggested by \citet{Ostriker2001} has been usually used to obtain magnetic field strengths of molecular clouds or cores with ordered field lines  (e.g., \citealt{Crutcher2004}; \citealt{Crutcher2012}).

The POL-2 polarimeter on the James Clerk Maxwell Telescope (JCMT) has been utilized for polarization observations in low- and high-mass star-forming regions. POL-2 has shown much better sensitivity than the previous polarimeter, SCUPOL (\citealt{Matthews2009}; \citealt{Friberg2016}). One of the JCMT large programs is the B-fields In STar-forming Region Observations (BISTRO) Survey which is aiming to study polarization properties and magnetic fields in star-forming clouds or cores \citep{Ward_Thompson2017}. The original BISTRO survey targeted nearby molecular clouds in the Gould Belt at distances around 130-450 pc. An extension of the BISTRO Survey (BISTRO-2) targeted high-mass star-forming regions located at distances $>$ 1 kpc. The third BISTRO survey (BISTRO-3) recently approved is aiming to study the magnetic fields in diverse evolutionary stages of star-forming regions, e.g., prestellar cores and massive clouds. The results of the BISTRO surveys have shown highly resolved magnetic fields and measurements of the field strengths in Orion A \citep{Pattle2017}, Oph-A \citep{Kwon2018}, Oph-B \citep{Soam2018}, Barnard 1 \citep{Coude2019}, Oph-C \citep{Liu2019}, IC 5146 \citep{Wang2019}, M16 \citep{Pattle2018}, NGC 6334 \citep{Arzoumanian2020}; and several other regions are under investigation.

Orion A is a nearby well studied high-mass star-forming region located at $\sim$400 pc (\citealt{Menten2007}; \citealt{Kounkel2017}). Orion A contains the OMC-1 region which has two clumps, BN-KL and South (S). Magnetic field strengths in the region have been estimated several times using the DCF method. The magnetic field strength over the entire OMC-1 region was estimated to have an average value of 0.76 mG by \citet{Houde2009} considering the effect of turbulent correlation length in measurement of polarization angle dispersion using polarization data obtained by the SHARP polarimeter on the Caltech Submillimeter Observatory (CSO). \citet{Pattle2017} calculated an average magnetic field of 6.6 $\pm$ 4.7 mG in the entire OMC-1 region using POL-2. They suggest an `unsharp masking' method to trace large scale field directions, which must be accounted for when estimating the angle dispersion for the DCF method. Recently, \citet{Chuss2019} estimated magnetic field strengths ranging from 0.26 to 1.01 mG in the BN-KL, Bar and HII regions in Orion A with the structure function suggested by \citet{Houde2009} using data taken by the High-resolution Airborne Wideband Camera-Plus (HAWC+) on the Stratospheric Observatory for Infrared Astronomy (SOFIA). Recently, \citet{Guerra2020} obtained maps of magnetic field strengths using the SOFIA data applying the structure function within a small circular sub-region at every pixel. The maximum value of the maps is about 2 mG. The field strength measurements in the OMC-1 region vary by an order of magnitude across the scale of the cloud, which is caused by different analyzed size or wavelengths.

In this paper, we propose to estimate the distribution of magnetic field strengths in the OMC-1 region using the DCF method. Previous studies have simply obtained a mean magnetic field strength over a whole cloud or quite a large area of a molecular cloud using the DCF method despite the fact that the true magnetic field strengths at different positions within the same molecular cloud vary. This is partially due to the poor sensitivity of previous polarization observations, in which there are too few polarization measurements over the molecular cloud to derive field strengths in different sub-regions of the cloud.   
The improved sensitivity of the POL-2 instrument provides us with a high-resolution polarization map of the whole OMC-1 region. We measured a magnetic field strength in the plane of the sky using the DCF method in a small box, 40$^{''}$ $\times$ 40$^{''}$. By moving the box over the region, we obtained the magnetic field strength distribution across the OMC-1 region.
Our method is a simple application of the DCF method to many sub-regions of a single large star-forming region assuming that there are enough polarization measurements in each sub-region to allow this approach. Our method can be easily applied to new millimeter and sub-millimeter polarization observations using POL-2 and other polarimeters such as the Balloon-borne Large-Aperture Submillimeter Telescope for polarization (BLAST-pol), the HAWC+ on the SOFIA and the Atacama Large Millimeter/submillimeter Array (ALMA).  

Our paper is organized as follows. In section \ref{sec:obs}, we describe the observed data towards the OMC-1 region and the reduced maps. We introduce our new application of the DCF method to measure the magnetic field strength distribution in section \ref{sec:meth}. Discussions are presented in section \ref{sec:disc}. We summarize our results in section \ref{sec:summ}.

\section{Observations and Data Reduction} \label{sec:obs}

In order to estimate magnetic field strengths using the DCF method in the OMC-1 region, we used two data sets taken using the JCMT - polarized dust continuum emission and spectral lines. The JCMT is located near the summit of Maunakea, Hawaii. It has a main instrument suite consisting of the SCUBA-2 camera (Submillimetre Common-User Bolometer Array 2; \citealt{Holland2013}), its associated polarimeter POL-2 (\citealt{Friberg2016}) and the HARP spectrometer (Heterodyne Array Receiver Program; \citealt{Buckle2009}). Polarization emission and C$^{18}$O spectral lines in the OMC-1 region have been obtained from these instruments. In this section, we show these two observational data sets, and explain how we reduce them. 

\subsection{SCUBA-2/POL-2 observations}\label{subsec:pol}

\begin{figure*}[ht!]
\epsscale{1.1}
\plotone{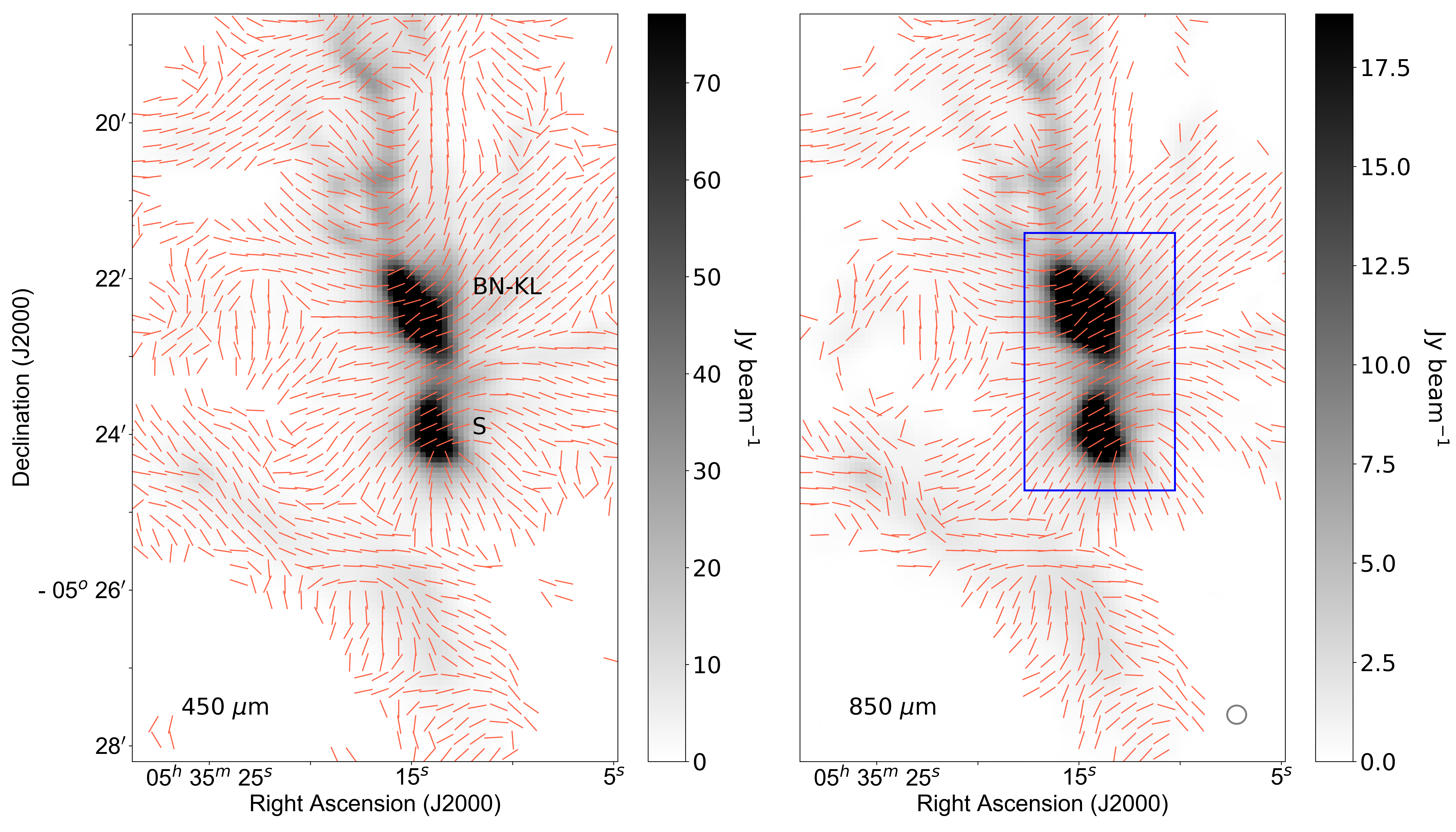}
\caption{Maps of magnetic field orientations where the originally observed polarization segments are rotated by 90 degrees of the OMC-1 region at 450 $\mu$m (left panel) and 850 $\mu$m (right panel). The segments are scaled to a uniform length for clarity The background images are total intensity (Stokes I) maps of the OMC-1 region at both wavelengths. Both show two bright clumps: the upper clump is BN-KL and the lower clump is S, and marked in the left panel. The beam size at 450 $\mu$m is 9.6 $\arcsec$. The map at 450 $\mu$m is convolved to have the same beam size of the map at 850 $\mu$m, 14.1$''$. The beam size is shown in the lower right corner of the right panel. the The polarization segments are selected using the criteria $I$/$\delta I \geq$ 10, $p$/$\delta p \geq$ 3 and $p$ $<$ 15\%, where $I$ and $p$ are total intensity and polarization fraction, respectively, and  $\delta I$ and $\delta p$ are the uncertainties of $I$ and $p$, respectively. The original 4$^{\prime \prime}$-sized polarization pixels are binned to 12$''$. Our analyzed region in this paper is shown as the blue box in the right panel. \label{fig:pol}}
\end{figure*}

Observations of total intensity and polarized dust continuum from the OMC-1 region are obtained by inserting the POL-2 polarimeter \citep{Friberg2016} into the light path of the SCUBA-2 camera. The polarization observations were performed with the POL-2 DAISY observing mode \citep{Friberg2016} at 450 $\mu$m and 850 $\mu$m, simultaneously. The central $3'$ diameter of the obtained map show a high signal-to-noise ratio and the noise increases from this circle to the edge of the map. The average scan speed is $4''$ s$^{-1}$ and the rotational speed of the half-wave plate is 8 Hz. 

The polarized emission in the OMC-1 region was obtained as part of the BISTRO large program (project code M16BL004; \citealt{Ward_Thompson2017}) and the POL-2 commissioning project (project code M15BEC02, \citealt{Friberg2016}, \citealt{Friberg2018}; Bastien et al. in prep.). In the BISTRO program, the OMC-1 region  was observed 21 times between January 11, 2016 and January 24, 2016 in Band 1 or Band 2 weather conditions, $\tau _{225 \text{GHz}} < 0.05$ or $0.05  \leq \tau _{225 \text{GHz}} < 0.08$, where $\tau_{225 \text{GHz}}$ is the atmospheric opacity at 225 GHz. The total on-source time is 14 hours. These data were first published by \citet{Ward_Thompson2017}. We excluded one of these datasets because the observation was incomplete.
A further set of 15 observations of the OMC-1 region were made between December 8, 2017 and January 10, 2018 as part of the POL-2 450 $\mu$m commissioning campaign.  As part of the effort to characterize the instrumental polarization at 450 $\mu$m, these observations were made without the JCMT Gore Tex in place.
Most of these observations were obtained in Band 1 weather condition ($\tau _{225 \text{GHz}} < 0.05$) with a total on-source time of 7.9 hours. We publish these observations of the commissioning campaign and the polarization data of the two projects at 450 $\mu$m for the first time. A total of 35 data sets at 450 $\mu$m and 850 $\mu$m from these two projects were used in this paper.  
 
 We combined all 35 data sets of the OMC-1 region and reduced them using the \textit{pol2map} routine in the Sub-Millimetre User Reduction Facility (SMURF) package of Starlink software \citep{Jenness2013} with the latest Instrumental Polarization (IP) model, 'August 2019'\footnote{\url{https://www.eaobservatory.org/jcmt/2019/08/new-ip-models-for-pol2-data/}}. There are three main stages of the POL-2 data reduction process. Polarized light is represented by the Stokes parameters, $I$, $Q$ and $U$. First, the raw time-series data is converted into Stokes $I$, $Q$ and $U$ time-streams using the \textit{calcqu} command. Then, the \textit{makemap} routine is used to make an initial reference Stokes $I$ map from each dataset, gridded onto $4''$ pixels. A co-added total intensity map is created from all of these initial Stokes $I$ maps. Next, \textit{pol2map} generates two `masks', named ASTMASK and PCAMASK, from the co-added total intensity map. The ASTMASK is a fixed signal-to-noise-based mask, which is used to define regions of astrophysical emission in the co-added total intensity map. The PCAMASK defines source regions in order to exclude the regions containing emission when making a background model with \textit{makemap}. See \citet{Mairs2015} for a detailed discussion of the role of masking in SCUBA-2 data reduction. \textit{pol2map} makes an improved Stokes $I$ map of each data set by using these fixed masks, increasing the number of principal component analysis (PCA) components over the first reduction, and using the \textit{skyloop} implementation of \textit{makemap}, wherein each iteration of the mapmaker is performed on each of the observations in turn, rather than each observation being reduced consecutively. The final Stokes $I$ map is obtained by co-adding the improved Stokes $I$ maps. \textit{pol2map} uses the same set of parameters and masks to create the Stokes $Q$ and $U$ maps of each dataset, which are then co-added. The variance maps of the co-added final Stokes $I$, $Q$ and $U$ maps are calculated from the variance in the values of the Stokes $I$, $Q$ and $U$ maps of all datasets. A polarization vector catalog is created by \textit{pol2stack} in SMURF using Stokes $I$, $Q$, $U$ and their variance maps. See \citet{Parsons2018} for a detailed description of POL-2 data reduction. 
 
 \begin{figure}[htb!]
\epsscale{1.0}
\plotone{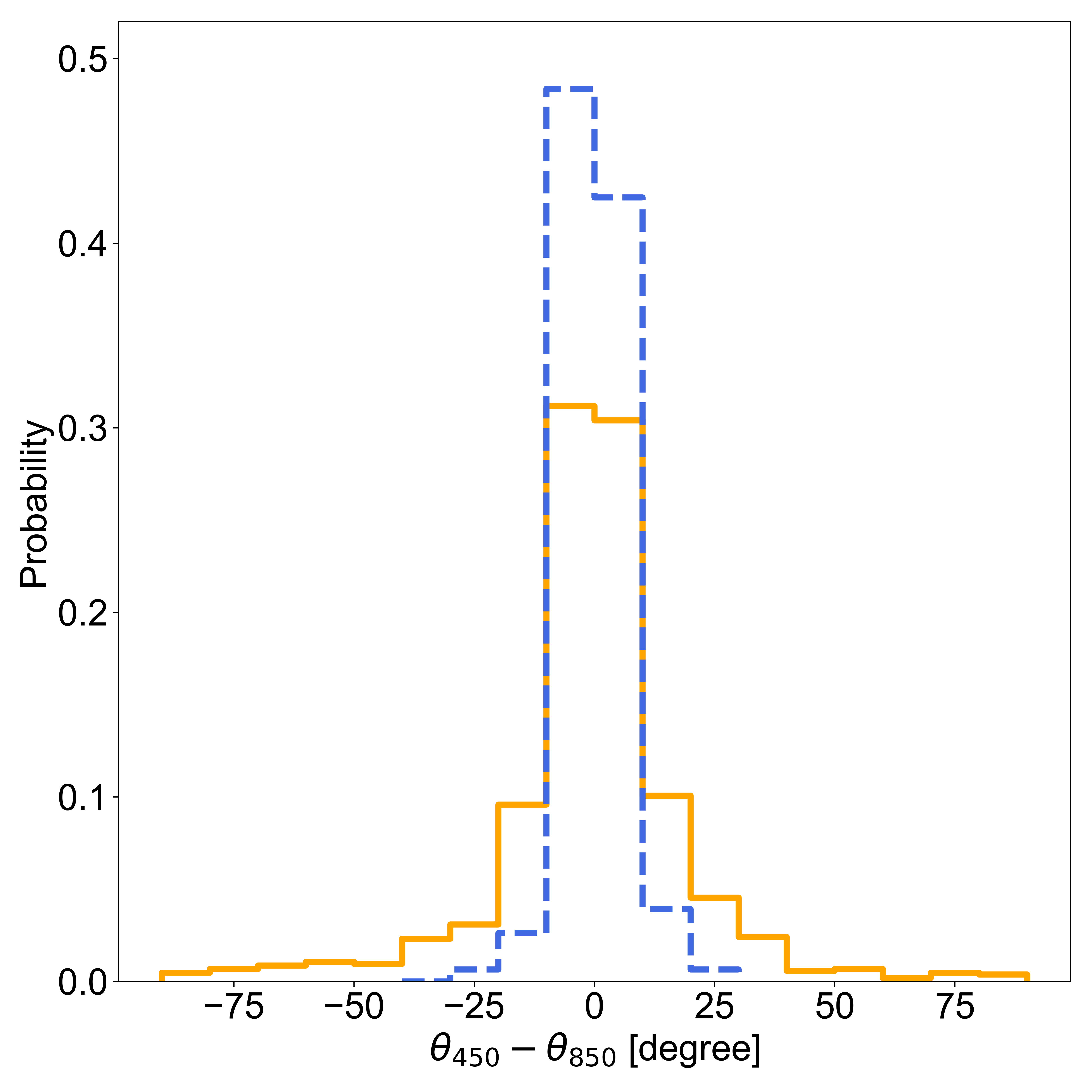}
\caption{Histograms of differences of all polarization angles in the Figure \ref{fig:pol} (orange colors) and the angles inside the blue box of the figure (blue colors) at 450 $\mu$m and 850 $\mu$m. \label{fig:polang}}
\end{figure}

 The final Stokes $I$, $Q$ and $U$ maps are given in units of pW. We converted these maps to units of Jy beam$^{-1}$ applying flux conversion factors of 962 and 725 Jy pW$^{-1}$ beam$^{-1}$ at 450 $\mu$m and 850 $\mu$m (\citealt{Dempsey2013};  \citealt{Friberg2018}). The root mean square (rms) noises of the $I$, $Q$ and $U$ maps are obtained by calculating the square root of the mean variance within an area in the eastern part of the OMC-1 region, known to be an ionized region. The rms noises of Stokes $I$ are 28.8 and 6 mJy beam$^{-1}$ at 450 $\mu$m and 850 $\mu$m, respectively, while those of Stokes $Q$ are 25 and 3.9 mJy beam$^{-1}$, and those of Stokes $U$ are 24.1 and 3.8 mJy beam$^{-1}$. Our polarization data at 850 $\mu$m contain not only BISTRO data but also additional polarization data of the OMC-1 region of the POL-2 commissioning project. We thus obtained a 23\% lower rms noise in Stokes $I$ at 850 $\mu$m than that previously obtained using  BISTRO survey data alone (e.g., \citealt{Ward_Thompson2017}; \citealt{Pattle2017}).

 \begin{figure*}[htb!]
\epsscale{0.7}
\plotone{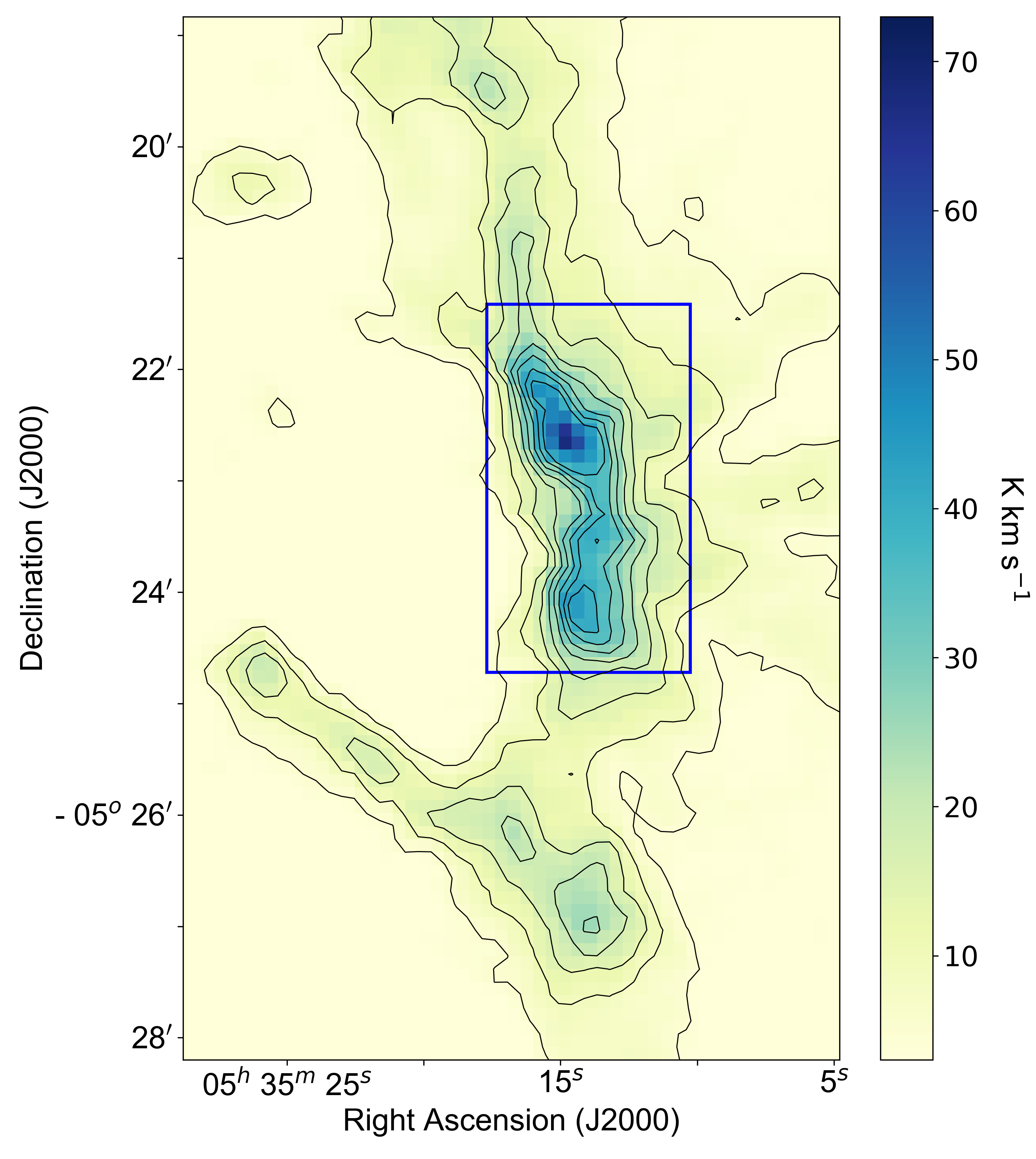}
\caption{JCMT HARP C$^{18}$O ($J$ = 3-2) integrated intensity map. Emission is integrated over the local standard of rest (LSR) velocity range -10 km s$^{-1}$ to 25 km s$^{-1}$. Contours run from 5 K km s$^{-1}$ to 40 K km s$^{-1}$ in steps of 5 K km s$^{-1}$. The blue box is the same as defined in Figure \ref{fig:pol}. \label{fig:c18o}}
\end{figure*}

Figure \ref{fig:pol} shows magnetic field orientations of the OMC-1 region at 450 $\mu$m and 850 $\mu$m. We convolved the data at 450$\mu$m to have the same beam size of 850 $\mu$m data, 14.1$^{\prime \prime}$, using \textit{smooth450} command in \textit{pol2map} routine. The background images in both panels of the figure represent total intensities at two wavelengths which show quite similar distributions. Each map contains two main clumps: the upper clump is BN-KL and the lower clump is S. The Orion Bar structure is shown in the south-east side of each panel. The magnetic field orientations are represented by rotating the polarization segments by 90 degrees. We binned the original 4$''$ pixels to 12$''$, which is close to the effective beam size of JCMT at 850 $\mu$m in Figure \ref{fig:pol} to avoid too many polarization segments. When analyzing this region, the original 4$''$ pixels are gridded to 8$''$ at 850 $\mu$m, which approximately corresponds to the Nyquist sampling interval of the JCMT at 850 $\mu$m. The overall distribution of magnetic field orientations in the OMC-1 region shows a hourglass morphology which is consistent with previous polarization observations at far-infrared and sub-millimeter (e.g., \citealt{Coppin2000}; \citealt{Ward_Thompson2017}; \citealt{Chuss2019}). The blue box of the right panel represents our analyzed region in next sections, which contains two clumps. We focus on the region due to high signal to noise ratio of polarization data, $p/\delta p >20$ and ordered structure of polarization segments.

We do not show the polarization fraction of each segment in Figure \ref{fig:pol} because of the following reason. In a previous BISTRO paper, \citet{Ward_Thompson2017} show that the polarization fractions in the very central region of the BN-KL clump decrease as increasing total intensity. They suggest the depolarization is caused by magnetic field lines tangled in the dense core. The dust polarization emission in the core with complex and small-scale magnetic field geometry is averaged within an antenna beam, so the polarization fractions at the core could be decreased. Another possibility of the depolarization at the core is due to the inefficiency of the dust alignment by radiation. \citet{Andersson2015} expect dust grains are less efficiently aligned with magnetic field lines at a high optical depth. Because of the two possibilities, our measured polarization angle dispersions at the core might be over-estimated in the small central region of the BN-KL clump. The over-estimate can finally affect the magnetic field strength and the mass-to-flux ratio at the clump.

Figure \ref{fig:polang} shows the histograms of the differences of all polarization angles in Figure \ref{fig:pol} and the angles inside the blue box in the figure at 450 $\mu$m and 850 $\mu$m. Polarization angles are measured from the north to east along the counter-clock-wise direction. A positive value of the difference of two polarization angels means that a polarization segment at 450 $\mu$m is located at a counter-clock-wise direction from a segment at 850 $\mu$m. 
Because of the 180-degree ambiguity of a polarization angle, the angle difference of two polarization segments is less than 90 degrees. The 85\% of angle differences of all polarization angles is in the range of -25 to +25 degrees (orange histrogram in Figure \ref{fig:pol}). More detailed analysis of the differences of polarization angles and angle dispersions is given in Appendix \ref{sec:app0}

\subsection{HARP observations}
 
 C$^{18}$O observations of the OMC-1 region were taken as part of the JCMT Gould Belt Survey (GBS; \citealt{Ward_Thompson2007}; \citealt{Salji2015a}; \citealt{Mairs2016}). These observations were made from February 2009 to October 2010 with the HARP instrument  tuned to the C$^{18}$O ($J$ = 3-2) frequency of 329.278 GHz \citep{Buckle2012}. The system temperatures varied from 225 to 689 K over an atmospheric opacity range of 0.03-0.07 at 225 GHz. The data were reduced using the ORAC Data Reduction (ORAC-DR) pipeline and the Kernel Application Package (KAPPA) \citep{Currie2008} in Starlink by \citet{Buckle2012}. We used the reduced C$^{18}$O map gridded to a pixel size of 8$''$. The C$^{18}$O integrated intensity map is presented in Figure \ref{fig:c18o} which is consistent with total intensity images in both panels of Figure \ref{fig:pol}. It shows the two main clumps and the Orion Bar structure shown in Figure \ref{fig:pol}

\section{Methods and Results \label{sec:meth}}

The DCF method is widely used for obtaining magnetic field strengths in the interstellar medium (ISM) from polarization observations. The magnetic field is flux-frozen with the gas such that distortion in the field morphology is due to small-scale non-thermal motions. The underlying assumption of the DCF method is that the distortion of magnetic field lines by turbulence is reflected into the dispersion of polarization angles. To measure the dispersion caused by turbulence, it is necessary to estimate the large-scale magnetic field structure. The magnetic field strength in the plane of the sky is estimated by measuring the volume density, the velocity dispersion and the dispersion of polarization angles. \citet{Crutcher2004} suggests a simple formula, 

 \begin{equation} 
B_{\text{pos}} = Q\sqrt{4\pi\rho}\frac{\sigma_{v}}{\sigma_{\theta}} \approx 9.3\sqrt{n(\text{H}_2)}\frac{\Delta V}{\sigma_\theta},
\label{eq:cf}
 \end{equation}
where $B_{\text{pos}}$ is magnetic field strength in the plane of the sky in units of $\mu$G; $Q$ is a correction factor which is suggested as 0.5 by \citet{Ostriker2001}; $\rho$ is the mean volume density in g cm$^{-3}$; $\sigma_{v}$ is the non-thermal velocity dispersion of the gas in km s$^{-1}$; $\sigma_\theta$ is the dispersion of polarization angles in degrees; $\rho=\mu m_{\text{H}} n(\text{H}_2)$, where $\mu$=2.8 is the mean molecular weight per particle by assuming 10\% of total gas number is helium \citep{Kauffmann2008} and $m_{\text{H}}$ is the mass of a hydrogen atom; $n(\text{H}_2)$ is the volume density of molecular hydrogen in units of cm$^{-3}$; $\Delta V$ is the full width at half maximum (FWHM) of the non-thermal component of a spectral line in units of km s$^{-1}$.

A dispersion of polarization angles is often measured as a standard deviation of the angles assuming that an underlying magnetic field is uniform with a direction equal to the mean orientation of the polarization segments over quite a large area of, or the whole area of, a molecular cloud or core (e.g., \citealt{Kirk2006}; \citealt{Curran2007}; \citealt{Cortes2016}; \citealt{Choudhury2019}). Other methods include a non-uniform magnetic field model to fit the overall shape of polarization segments \citep{Girart2009}, a two-point correlation function to determine the field structure function (e.g., \citealt{Hildebrand2009}; \citealt{Houde2009}; \citealt{Poidevin2010}; \citealt{Chuss2019}), and a spatial-filter to estimate the underlying field morphology \citep{Pillai2015}. \citet{Pattle2017} applied an `unsharp masking' method that uses a moving average of polarization angles with a 36$^{\prime \prime} \times$36$^{\prime \prime}$ subregion of Orion A to get the large-scale mean magnetic field. They measured  the angle difference between the mean directions and original local field directions at every position of the Orion A region. They calculated a polarization angle dispersion across the whole region by taking the standard deviation of the angle differences in the area over which the measurement error in angles is less than 2 degrees. Recently, \citet{Guerra2020} applied a two-point correlation function to the polarization maps of the Orion A region obtained by SOFIA to obtain the maps of magnetic field strengths in the region.

Here, we present a new application of the DCF method, which is an extension of the unsharp masking approach, to estimate the magnetic field strength distribution in the OMC 1 region. We obtain the distributions of the volume density from dust emission, velocity dispersion from C$^{18}$O spectral lines, and angle dispersion from polarization observations. We then estimate magnetic field strengths using Equation (\ref{eq:cf}) over the region. The detail procedures are explained in the following subsections.
 
  \subsection{Polarization angle dispersion\label{subsec:angdisp}}

It is necessary to estimate a mean magnetic field direction to accurately determine the turbulent dispersion of magnetic field lines. Since magnetic field lines are mainly distorted in a molecular cloud by both turbulence and gravity, it is not easy to determine a local mean field orientation from polarization observations. We estimate a mean field orientation in a small moving box in the OMC-1 region to trace its underlying morphology. By repeating the estimation over the region, we obtain the distribution of mean field orientations over the OMC-1 region. The collection of mean field directions will trace the large-scale variation of field lines over the region. This process is similar to the method suggested by \citet{Pattle2017}. \citet{Pattle2017} found good agreement between this moving box average and the true field direction using Monte Carlo simulations. We then estimate an angle dispersion from the differences of the original angles and the estimated mean angles in a box. We calculate angle dispersion as a root mean squared of the angle differences in the box. The further explanations are in Appendix \ref{sec:appendixA}. 

\begin{figure*}[h!]
\epsscale{1.0}
\plotone{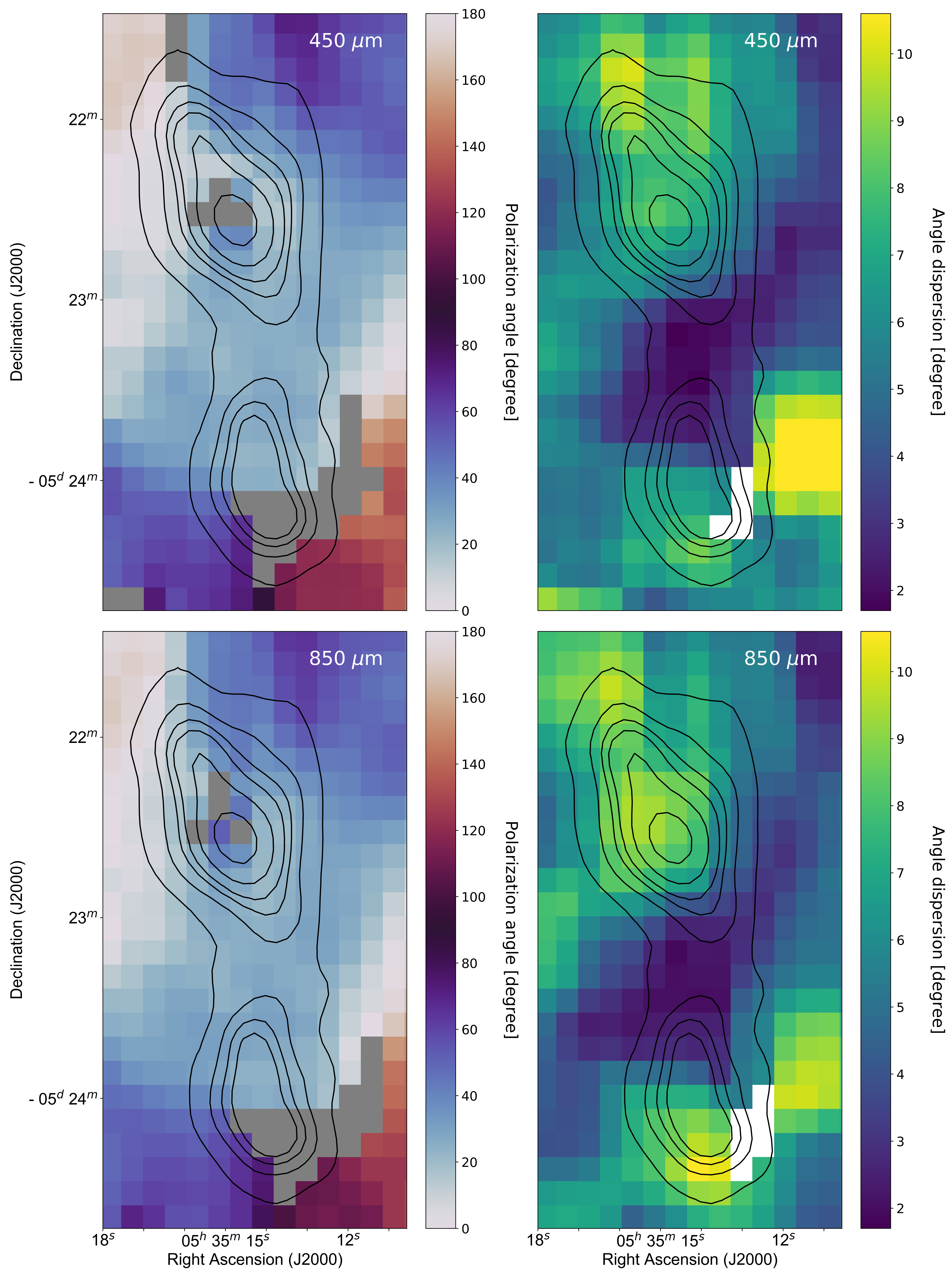}
\caption{A polarization angle distribution in the OMC-1 region at 450 $\mu$m (upper left panel) and 850 $\mu$m (lower left panel). Gray pixels whose curvature radii are smaller than 40$''$ are discarded when the calculation of the angle dispersion is made. The angle dispersions in the same region at 450 $\mu$m (upper right panel) and 850 $\mu$m (lower right panel) are estimated using a moving 5$\times$5 pixel box in case the total number of valid pixels is larger than the half of pixels in the box. Contours in both maps have equal Stokes $I$ at 850 $\mu$m, whose intensities are 6, 12, 18, 24, 40 and 80 Jy beam$^{-1}$. \label{fig:f6}}
\end{figure*}

Our method is similar to the `unsharp masking' method from \citet{Pattle2017}, but there are three main differences. First, we use 8$^{\prime \prime}$ pixels which is about Nyquist sampling instead of 12$^{\prime \prime}$ to minimize information loss and show well resolved polarization image. \citet{Pattle2017} calculated a mean value of nine angles in a 3$\times$3 pixel box. We calculate the mean angle in the box using the mean Stokes $Q$ and $U$ values, $\overline{Q}$ and $\overline{U}$, at 25 pixels in a 5$\times$5 pixel box. We move the box and repeat the calculation of mean angle over the OMC-1 region. Finally, we calculate an angle dispersion in the box as the root mean squared of the angle differences between the observed angles ($\theta_i$) and estimated mean angles ($\overline{\theta}$) for all 25 pixels in a box, $\sqrt{\sum_{i=1}^{25}(\theta_i-\overline{\theta})^2/25}$.  However, \citet{Pattle2017} calculated the standard deviation of angle differences ($\Delta \theta=\theta-\overline{\theta}$), $\sqrt{\sum_{i=1}^{N}(\Delta\theta_i - \overline{\Delta\theta}_i)^2/N}$, where $\overline{\Delta\theta}$ is the mean value of angle differences and $N$ is the total number of angle differences. We think that subtracting the mean value of angle differences could underestimate the angle dispersion in the box. \cite{Pattle2017} estimated a mean angle dispersion for the entire OMC-1 region, but we obtain angle dispersions in the region by moving the box. Our method can show highly resolved distribution of magnetic field strengths for the entire OMC-1 region compared to \citet{Pattle2017} method. 

The angle dispersion is dependent upon the box size, so we have to determine an appropriate box size. The box size has to be smaller than any 
large-scale non-turbulent feature imprinted on the magnetic field morphology. The field curvature can be adopted as a measure for this. We calculated a radius of curvature which is defined as the radius of a circle at a point on a curve which best approximates the curve at the point. We measure the radii of curvature over the OMC-1 region using polarization data in Appendix \ref{sec:appendixB}. Most of the radii in the region are larger than 40$\arcsec$, so we can choose a box with a side length of 40$\arcsec$ or less, a 5$\times$5 or a 3$\times$3 pixel box (pixel size = 8$\arcsec$). When the radii of curvature are larger enough than the box sizes, angle dispersions estimated in the 3$\times$3 and 5$\times$5 pixel boxes are similar. Since 25 pixels would have much better statistics for a mean and standard deviation than 9 pixels, we choose the 5$\times$5 pixel box as the box size for getting an angle dispersion instead of the 3$\times$3 pixel box. The details are explained in Appendix \ref{sec:appendixB}.  

Figure \ref{fig:f6} shows the distributions of polarization angles and angle dispersions at 450 $\mu$m and 850 $\mu$m obtained using a 5$\times$5 pixel box. The maps at two wavelengths are similar and the differences of the maps are described in Appendix \ref{sec:app0}. We are interested in a dense region containing the two clumps and having intensity larger than 6 Jy beam$^{-1}$ in Stokes $I$ map at 850 $\mu$m, which is inside the lowest contour line in each panel. In this region inside the contour line, the signal to noise ratio (S/N) of C$^{18}$O spectral line is larger than 10 and $p/\delta p$ is larger than 20. We obtain distributions of all the quantities required in the DCF method inside this contour level due to these high S/N of both of spectral line and polarization emission. The radii of curvature at gray pixels in the figure are smaller than 40$''$ which is a side length of the box. When we calculate an angle dispersion of polarization segments at  pixels whose curvature radii are smaller than box size, the angle dispersion is overestimated (Appendix \ref{sec:appendixB}). Because of this reason, we discard those pixels when we calculate angle dispersions. We calculate an angle dispersion at the center pixel of a box if the remaining pixels are larger than 50\% of the total number of pixels in the box.
 The ranges of angle dispersions inside the lowest contour line range from 1.9 to 10.6 degrees at 850 $\mu$m and from 1.7 to 10.1 degrees at 450 $\mu$m. Their mean angle dispersions over the analyzed region at 450 $\mu$m and 850 $\mu$m are 6.3 and 6 degrees which are larger than the mean angle dispersion estimated by \citet{Pattle2017}, 4 degrees. Since all angle dispersions inside the lowest contour level are less than 25 degrees, we can apply the DCF method modified by \citet{Ostriker2001} to estimate magnetic field strengths in the OMC-1 region.

\subsection{Volume density\label{subsec:col}}
We estimate the volume density of molecular hydrogen within the OMC-1 region from Stokes $I$ maps gridded to 8$''$ at 450 $\mu$m and 850 $\mu$m. Firstly, we correct CO contamination to the dust continuum at 850 $\mu$m. Flux density measured using the SCUBA-2 850 $\mu$m wide-band filter can include a contribution from the $^{12}$CO ($J$ = 3-2) line if there is a bright outflow (e.g., \citealt{Johnstone2003}; \citealt{Drabek2012}). Previous studies show that CO contamination is mostly small ($<$5\%), but toward bright outflows the line emission can contribute up to $\sim$15-20\% of the flux density at 850 $\mu$m (see also,  \citealt{Sadavoy2013}; \citealt{Coude2016}). The CO contamination is subtracted in Stokes $I$, so it does not affect the Stokes $Q$ and $U$. Because of this reason, a polarization fraction can be increased, but a polarization angle will not be changed. Due to the explosive outflow in the BN-KL clump, the OMC-1 region shows strong $^{12}$CO ($J$ = 3-2) emission. We used $^{12}$CO ($J$ = 3-2) data taken as part of the JCMT GBS \citep{Buckle2012} and subtracted the $^{12}$CO emission from the dust continuum at 850 $\mu$m in \textit{pol2map} routine. The details are described SCUBA-2 Data Reduction Tutorial\footnote{\url{https://www.eaobservatory.org/jcmt/science/reductionanalysis-tutorials/scuba-2-dr-tutorial-5/}}. 

To measure the volume density from thermal dust emission, we need to estimate the dust temperature. We follow processes described in \citet{Salji2015a} which are also used by \citet{Pattle2017}. We use the intensity ratio of 450 $\mu$m and 850 $\mu$m data obtained by SCUBA-2/POL-2 observations (e.g., \citealt{Hatchell2013}) to estimate the dust temperature in the OMC-1 region. We convolve the 450 $\mu$m map to have the same beam size as the 850 $\mu$m map described in section \ref{subsec:pol} and measure the dust temperature in each pixel based on the intensity ratio,

 \begin{equation}
\frac{I_{850}}{I_{450}} =   \bigg(\frac{\nu_{850}}{\nu_{450}} \bigg) ^{3+\beta} \Bigg( \frac{e^{h\nu_{450}/kT}-1}{e^{h\nu_{850}/kT}-1} \Bigg),
\label{eq:temp}
 \end{equation}
 where $I_{850}$ and $I_{450}$ are intensities at 850 $\mu$m and 450 $\mu$m, respectively, and $\nu_{850}$ and $\nu_{450}$ are frequencies at 850 $\mu$m and 450 $\mu$m.

By assuming the dust emission from the OMC 1 region is optically thin and follows a modified blackbody distribution, we could estimate the column density as follows:

 \begin{eqnarray}
N(\text{H}_2) = \mu  m_{\text{H}}  \kappa(\nu) B_\nu(T)/I_\nu   \nonumber  \\
=\mu  m_{\text{H}} \kappa_{\nu _0}  \bigg(\frac{\nu}{\nu_0} \bigg) ^\beta B_\nu(T)/I_\nu,
\label{eq:bb}
 \end{eqnarray}
where  $N(\text{H}_2)$ is the column density of molecular hydrogen, $\kappa(\nu)$ is the dust opacity, $B_\nu(T)$ is the Planck function at dust temperature \textit{T} \citep{Hildebrand1983} and $I_\nu$ is the intensity at frequency $\nu$.
Dust opacity is determined by the dust opacity $\kappa_{\nu_0}$ at the rest frequency $\nu_0$ and the dust opacity spectral index $\beta$. We take $\kappa_{\nu_0}$ = 0.1 cm$^2$ at $\nu_0$ = 1000 GHz, take $\beta$ = 2 and assume a dust-to-gas mass ratio of 1:100 (\citealt{Beckwith1990}; \citealt{MotteAndr2001}; \citealt{Andre2010}).

 \begin{figure*}[thb!]
\epsscale{1}
\plotone{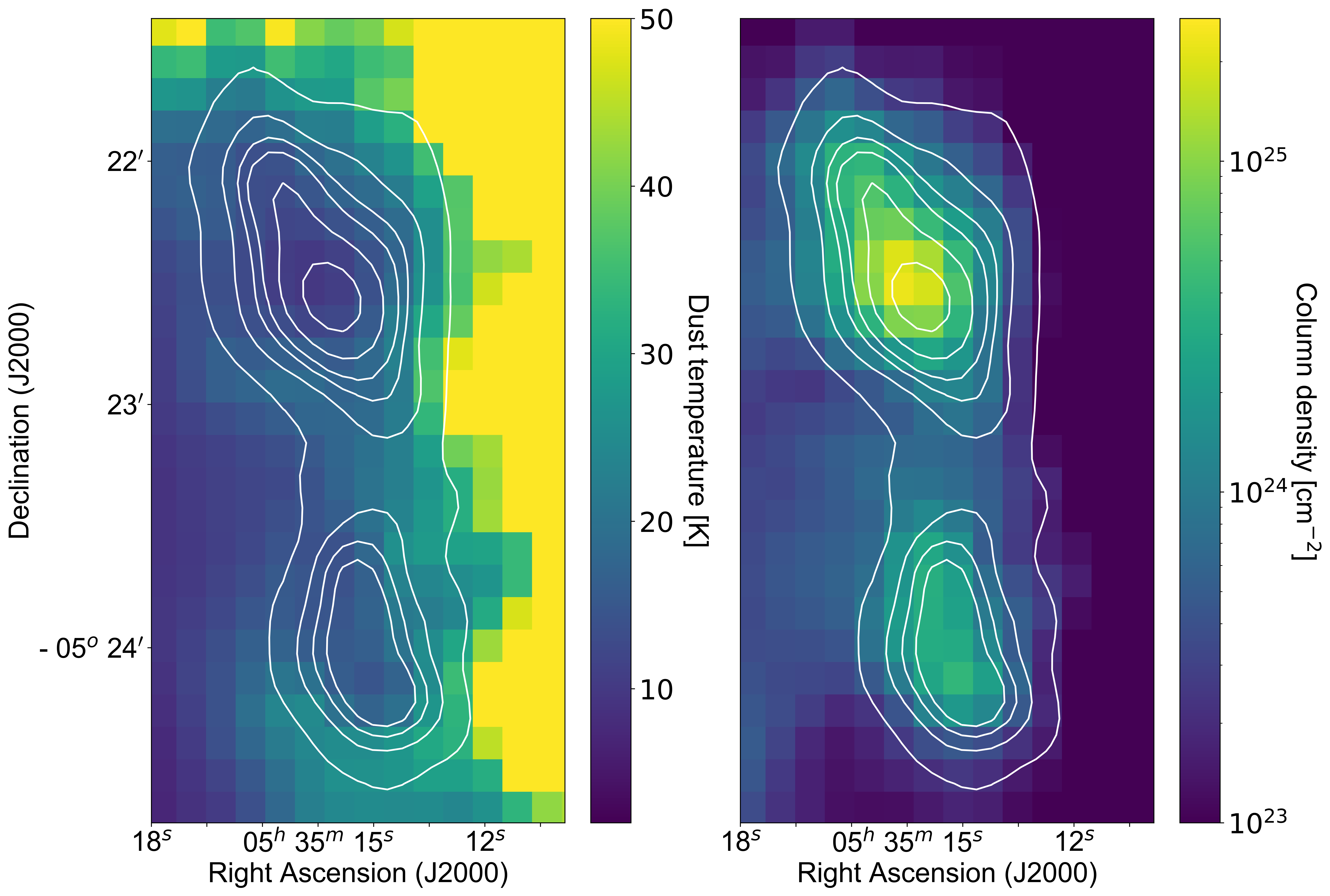}
\caption{The dust temperature map in the OMC-1 region estimated using Equation (\ref{eq:temp}) (left panel). The column density map obtained using Equation (\ref{eq:bb}) (right panel). Contours are the same as defined in Figure \ref{fig:f6} \label{fig:tempcol}}
\end{figure*}

The left panel of Figure \ref{fig:tempcol} shows the temperature distribution in the OMC-1 region. The intensities at 450 and 850 $\mu$m at temperatures higher than 50 K fall on the tail part of the Rayleigh-Jeans blackbody function. This is a situation where it is unreliable to get a temperature based on the two intensities at the tail. In fact, we put a constraint of a maximum temperature of 50 K when we calculate dust temperature from equation (\ref{eq:temp}). The inner region with the lowest contour level shows lower temperature than 50 K. The two main clumps are found to be at lower temperatures than other regions. We obtain a column density for each pixel from Equation (\ref{eq:bb}) using the dust temperature and the 850 $\mu$m intensity in the pixel (right panel of Figure \ref{fig:tempcol}). The range of column densities in the OMC-1 region is 8.5$\times$10$^{22}$-2.3$\times$10$^{25}$ cm$^{-2}$. It is consistent with previous results using dust continuum obtained by the GBS, 10$^{22}$-10$^{25}$ cm$^{-2}$ (\citealt{Salji2015a}; \citealt{Pattle2017}). The column density map clearly shows the two main clumps, and that BN-KL is denser than S. 

We assume the OMC-1 region is cuboid. Polarized dust emission we observed is an integrated polarized light along  a sight line. In principle the DCF method requires information of the three quantities in Equation (\ref{eq:cf}) in three-dimensional volume space. It is not possible to get the three-dimensional information from dust polarization and spectral line observations. In order to overcome this difficulty, researchers assume that integrated polarized emission comes from a specific depth along a sight line at which they assume an effective volume density. Then they obtain a magnetic field strength in the plane perpendicular to the sight line at the depth. In order to obtain an effective volume density in this study, we assume that the volume density in the OMC-1 region is proportional to a column density.  A simplest form of the proportionality is to assume a constant depth over the region that we are interested in. We determine the constant depth in the following way. \citet{Hacar2018} observed this region with  N$_2$H$^+$ ($J$=1-0) and  suggested that the volume density of the region is in the range of 10$^5$ -10$^6$ cm$^{-3}$. \citet{Teng2020} also estimated the volume density in the core region to be either 10$^7$ or 3$\times$10$^7$ cm$^{-3}$. The volume density is one of parameters of their non-LTE analysis to fit integrated intensity of N$_2$H$^+$ ($J$=3-2) spectral lines and the line ratio of  N$_2$H$^+$ ($J$=3-2) to N$_2$H$^+$ ($J$=1-0).  We choose the higher volume density to derive the constant depth.  We divide the column density peak of the BN-KL clump 2.3$\times$10$^{25}$ cm$^{-2}$, a higher end of the column density range obtained in the previous paragraph, by the volume density, 3$\times$10$^7$ cm$^{-3}$. An  estimated depth ($W$) ,$\sim$0.2 pc, is largely consistent with the typical width of filaments $\sim$0.1 pc estimated by \textit{Herschel} observations in nearby clouds like IC5146, Taurus or Polaris (\citealt{Arzoumanian2011}; \citealt{Li2012}; \citealt{Palmeirim2013}; \citealt{Andre2014}; \citealt{Arzoumanian2019}). \citet{Pattle2017} assumed the OMC-1 region is a cylindrical filament with radius $r$ = 0.09 pc and length $L$ = 0.35 pc and \citet{Chuss2019} used an uniform depth $\sim0.15$ pc based on \citet{Pattle2017}. 
It is $\sim$1.3 times smaller than our estimated width. 
 
The volume density is calculated using the following relation:
  \begin{equation}
n(\text{H}_2) = \frac{N(\text{H}_2)}{W}
\label{eq:vol_den}
 \end{equation}
$W$ is the depth, 0.2 pc, which is estimated above. The range of obtained volume densities is 1.4$\times$10$^5$-3.0$\times$10$^7$ cm$^{-3}$. 
 
\subsection{Velocity dispersion}

\begin{figure*}[thb!]
\epsscale{0.9}
\plottwo{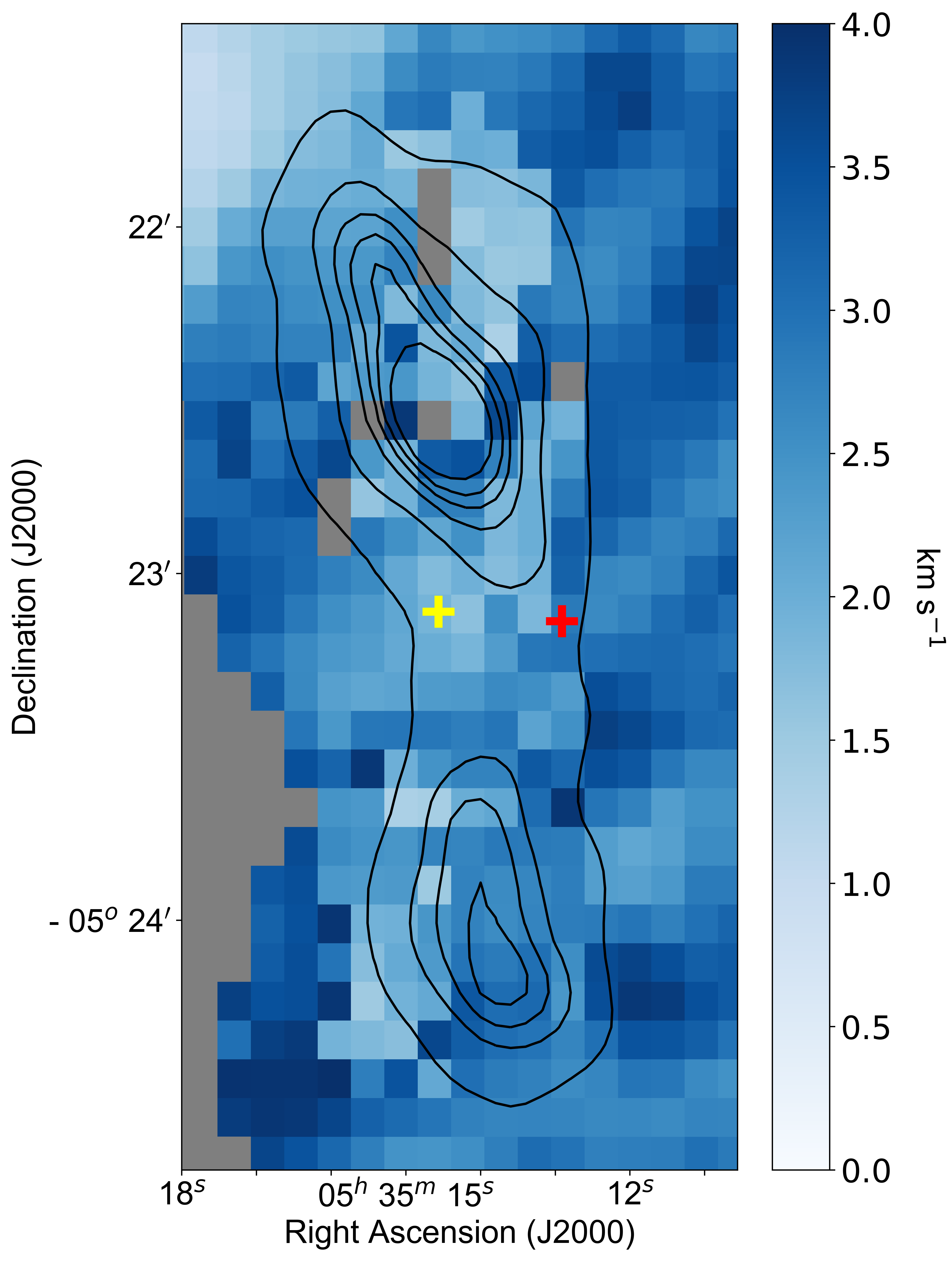}{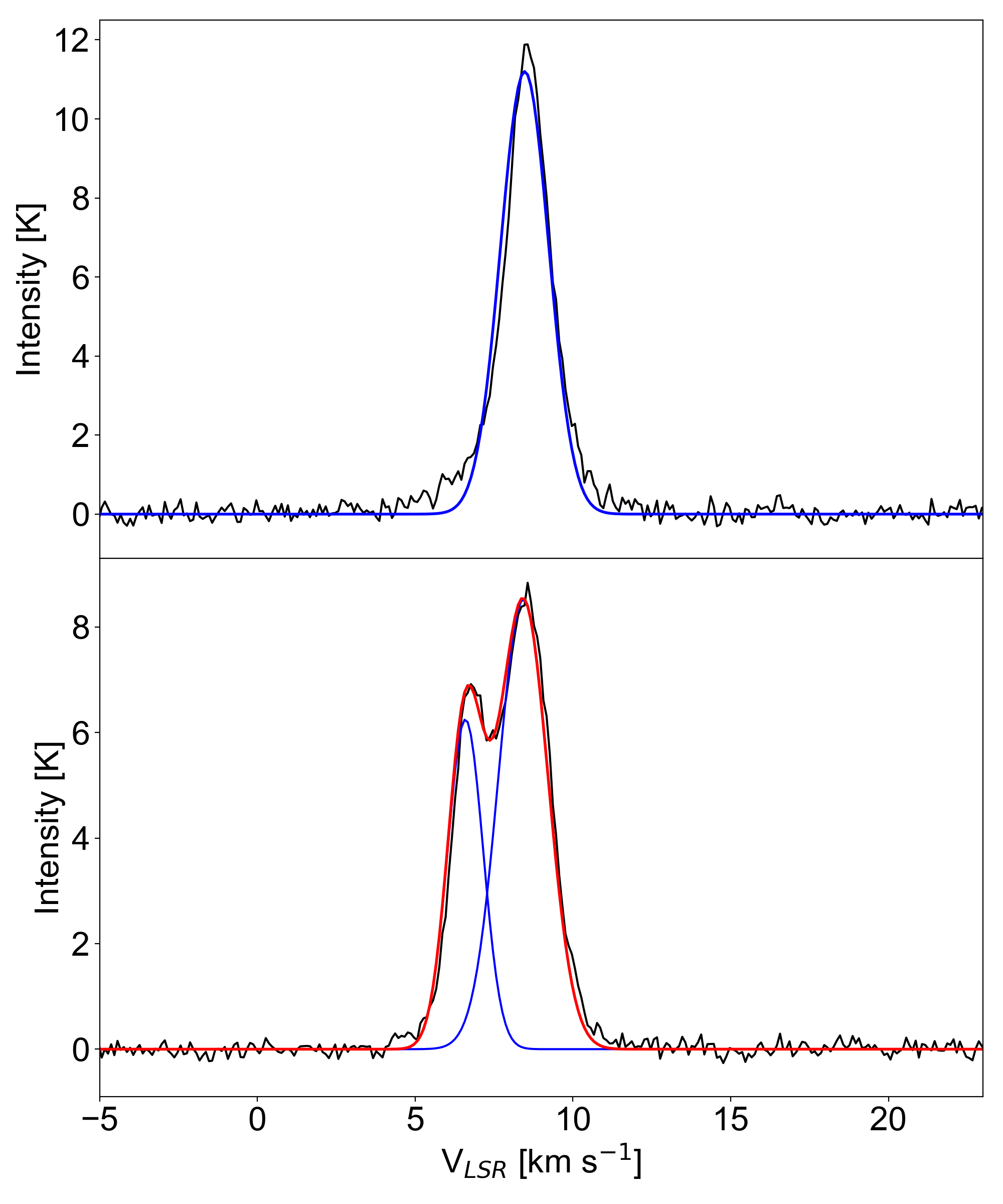}
\caption{The FWHM map of the non-thermal component of C$^{18}$O spectral lines in the OMC-1 region (left panel). Contours are the same as those defined in Figure \ref{fig:f6}. The peak value inside the lowest contour level is  3.9 km s$^{-1}$. The gray pixels contain spectral lines with unresolved blended lines. Examples of single and blended C$^{18}$O spectral lines (right panels). The upper right panel shows a single line at the location of a yellow cross shown in the left panel.  A fitting line with a single Gaussian component is shown with a blue line.  The lower right panel shows a blended line at the location of a red cross shown in the left panel.  A fitting line with two Gaussian components is shown with a red line.  
\label{fig:f5}}
\end{figure*}

 We use the C$^{18}$O ($J$ = 3-2) data from the HARP to measure the velocity dispersion. The HARP data have the same spatial resolution as SCUBA-2 at 850 um and were gridded to 8$\arcsec$ pixels to match the SCUBA-2/POL-2 observations. The integrated intensity map of the C$^{18}$O lines is similar to dust continuum (see Figure \ref{fig:c18o} and \ref{fig:pol}), so we assume C$^{18}$O traces a region emitting dust continuum at 850 $\mu$m. We fit a single Gaussian profile to the C$^{18}$O spectral line at each pixel using Continuum and Line Analysis Single-dish Software (CLASS) (\citealt{Pety2005}; \citealt{gildasteam2013}). Some pixels in the OMC-1 region contain double C$^{18}$O spectral lines having two different velocity components or unresolved blended lines. \citet{Pattle2017} excluded pixels showing double peaks or broad wings. We also exclude highly blended lines, but we fit C$^{18}$O spectral lines with double Gaussian profiles which can be resolved into two components. And then we select one profile of the two based on the following procedures. We first make a comparison of a dust continuum map of the OMC-1 region with each of C$^{18}$O channel maps in steps of 0.5 km s$^{-1}$ in the LSR velocity range of 6 to 9 km s$^{-1}$ made by \citet{Buckle2012}. We find that a channel map from 7.5 to 8 km s$^{-1}$ is well-matched with the continuum map. We finally choose one of the two blended components whose velocity is in this velocity range or close to the range. 
 
 We calculate a difference of the FWHM of a discarded line component form that of a selected line component. There are two velocity components at about 30\% of total pixels in the range delineated by the lowest contour level of the Figure \ref{fig:f5}. The mean value of the differences at the 30\% pixels is 0.3 km s$^{-1}$ and most of the differences are in a range from -0.75 km s$^{-1}$ to 0.5 km s$^{-1}$. The mean difference, 0.3 km s$^{-1}$, is 12.5\% of a mean FWHM of the selected velocity components in the region. So the estimated mean magnetic field strength has an uncertainty of about 12.5\% if the second velocity component is considered. We note that, since a volume density at a pixel is determined solely by dust emission at 450 $\mu$m and 850$\mu$m, the alternative choice of a velocity component doest not make any effect on the determination of the volume density at that pixel.


 We obtain a C$^{18}$O FWHM from a single Gaussian profile or a chosen Gaussian profile of a blended line. To estimate the non-thermal component, we subtract the thermal component from the measured C$^{18}$O FWHM,

\begin{equation}
\Delta V^2 = \Delta V_{\text{total}}^2 - \frac{ k T_k }{m_{\text{C}^{18}\text{O}}}\,8 \ln 2,
\label{eq:vel}
\end{equation} 
where $\Delta V$ is the FWHM of the non-thermal component, $\Delta V_{\text{total}}$ is the measured FWHM of the C$^{18}$O spectral line, $\sqrt{k T_k/m_{\text{C}^{18}\text{O}}}$ is the thermal velocity dispersion, $T_k$ is a kinetic temperature, and $m_{\text{C}^{18}\text{O}}$ is the mass of the C$^{18}$O molecule. We assume that an excitation temperature of C$^{18}$O is consistent with the kinetic temperature. \citet{Buckle2012} estimated the excitation temperature of C$^{18}$O spectral lines using the opacity of the CO isotopologue transitions estimated by the line-peak ratios of $^{13}$CO/C$^{18}$O, assuming a condition of local thermodynamic equilibrium (LTE). They estimated the mean values of the C$^{18}$O ($J$ = 3-2) excitation temperature in the OMC-1 region is 37.6 K \citep{Buckle2012}. We use this mean value as a gas temperature in Equation (\ref{eq:vel}). The mean FWHM of the thermal component is about 10.2\% of that of C$^{18}$O linewidth within the lowest contour level in Figure \ref{fig:f5}. The figure shows the distribution of non-thermal FWHMs in the OMC-1 region. The peak value is 3.9 km s$^{-1}$ inside the lowest contour level in the figure. The mean FWHM of non-thermal component within the lowest contour level is 2.4 km s$^{-1}$.

 We check whether the C$^{18}$O is depleted in the OMC-1 region or not by making a plot of  the integrated intensity of the C$^{18}$O as a function of the intensity of dust emission within the lowest contour level. The integrated intensity of the C$^{18}$O increases almost linearly as the intensity of dust emission at 850$\mu$m increases. Additionally, we compare the FWHM and integrated intensity of C$^{18}$O with these of N$_2$H$^+$, which is a well-known tracer of a dense region. \citet{Tatematsu2008} estimated line widths of 34 cores in the Orion A cloud using the N$_2$H$^+$ ($J$=1-0) spectral line with the Nobeyama 45m radio telescope. One of the cores is in our analyzed region. The line width of the N$_2$H$^+$ at the central position of the core is about 2.1 km s$^{-1}$. The FWHM of C$^{18}$O estimated by us at the same position is about 2.7 km s$^{-1}$, which is 28\% broader than the line width of N$_2$H$^+$. In dense regions containing cores, the velocity dispersion measured in C$^{18}$O is likely overestimated. While the integrated intensity of C$^{18}$O is well-matched with dust continuum, that of N$_2$H$^+$ obtained by \citet{Tatematsu2008} is not. The integrated intensity of N$_2$H$^+$ is shifted toward the western part of the BN-KL clump. Although the C$^{18}$O is not an ideal choice to trace dense regions, we believe that C$^{18}$O is the best choice to estimate velocity dispersions of the whole OMC-1 region.

\subsection{The distribution of magnetic field strengths}

\begin{figure*}[thb!]
\epsscale{1.0}
\plotone{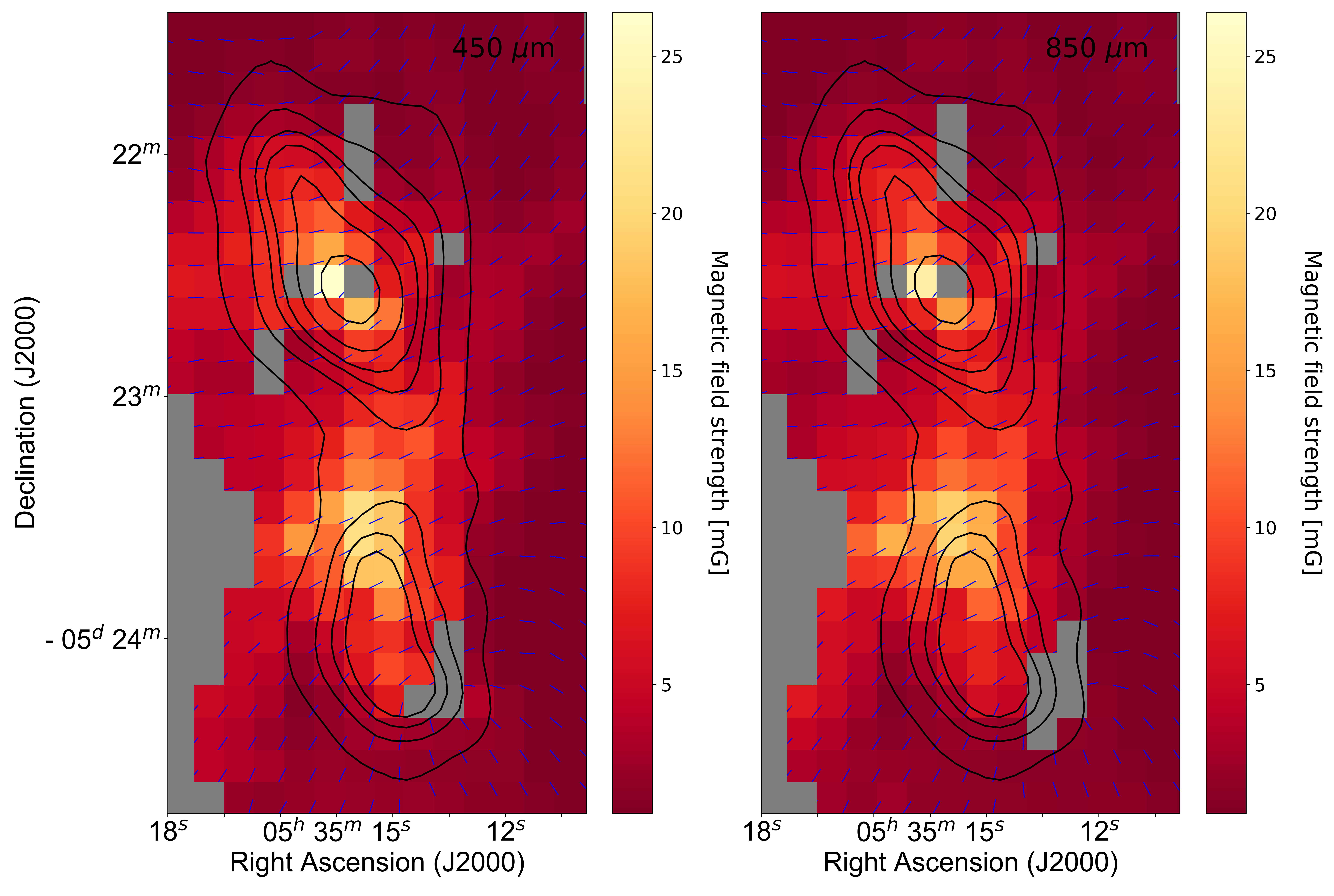}
\caption{The distribution of magnetic field strengths estimated using the DCF method in the OMC-1 region at 450 $\mu$m (left panel) and 850 $\mu$m (right panel). The blue segments are the orientations of magnetic field. Contours are the same as defined in Figure \ref{fig:f6}. In gray pixels, we could not get magnetic field strengths due to gray pixels in Figure \ref{fig:f5} and white pixels in the right panels of the Figure \ref{fig:f6}. Note that the polarization angle dispersion for the DCF method is obtained in a 40$''$ box and the box moves by 8$''$ over the region. \label{fig:mag}}
\end{figure*}

The distribution of magnetic field strength is obtained using three values of volume density, velocity dispersion and polarization angle dispersion. Three maps of column density, velocity dispersion and polarization angle dispersion obtained in previous sub-sections have the same resolution and  pixel size, 8$''$. The equatorial coordinates of the center position of each pixel in the column density map are the same with those in the angle dispersion map and different with those in the velocity dispersion map. We match coordinates of velocity dispersion with other two maps by interpolating values of velocity dispersion. We made the distributions of magnetic field strengths in the OMC-1 region at 450 $\mu$m and 850 $\mu$m (Figure \ref{fig:mag}) using Equation (\ref{eq:cf}) at each pixel.

 In Figure \ref{fig:mag}, the orientations of magnetic field at both wavelengths are represented as blue segments at both panels. At gray pixels in the figure, we could not obtain magnetic field strengths due to discarded values in velocity dispersion (gray pixels in Figure \ref{fig:f5}) or angle dispersion maps (white pixels in the right panels of Figure \ref{fig:f6}). The magnetic field strengths vary from 0.9 to 26.4 mG at 450 $\mu$m and from 0.8 to 24.4 mG at 850 $\mu$m inside the lowest contour level (6 Jy beam$^{-1}$) in both panels. The mean and median values are 6.6 $\pm$ 3 mG, 6 mG at 450 $\mu$m, and 6.2 $\pm$ 2.8 mG, and 5.3 mG at 850 $\mu$m.These uncertainites in the means are based solely on the range of inferred values over the region and not on the uncertainties in deriving those values. The magnetic field strengths at both wavelengths show good agreement within their uncertainties, so dust grains traced at both wavelengths can show similar polarization and magnetic field properties.

We find that the wider ranges of angle dispersions from 1.7 to 10.6 degrees and volume densities from 1.4$\times$10$^5$ to 3.0$\times$10$^7$ cm$^{-3}$ compared with the range of velocity dispersion, make wider the distribution of magnetic field strengths in the region. The non-thermal velocity dispersions have a relatively narrow range of values from 1.5 to 3.1 km s$^{-1}$, which suggests relatively similar non-thermal motions across the region. The narrow velocity range does not widen the distribution of magnetic field strengths. The uncertainties in our estimates of the volume density, velocity dispersion, polarization angle dispersion, and magnetic field strengths are discussed in the next section and detail discussions about the maps of magnetic field strengths are in section \ref{subsec:Bdisc}.

\subsection{Error analysis}

The fractional uncertainty in the magnetic field strength is expressed as 

\begin{equation}
\frac{\delta B_{\text{pos}}}{B_{\text{pos}}}=\sqrt{\bigg (\frac{1}{2}\frac{\delta n(\text{H}_2)}{n(\text{H}_2)} \bigg)^2+\bigg( \frac{\delta \Delta V}{\Delta V} \bigg ) ^2+ \bigg ( \frac{\delta \sigma_\theta}{\sigma_\theta} \bigg ) ^2},
\label{eq:db_b}
\end{equation}
where $\delta B_{pos}$ is the uncertainty in magnetic field strength in the plane of the sky, $\delta n(\text{H}_2)$ is the uncertainty in volume density, $\delta \Delta V$ is the uncertainty in FWHM and $\delta \sigma_\theta$ is the uncertainty in polarization angle dispersion.

\begin{figure*}[htb!]
\epsscale{1.0}
\plotone{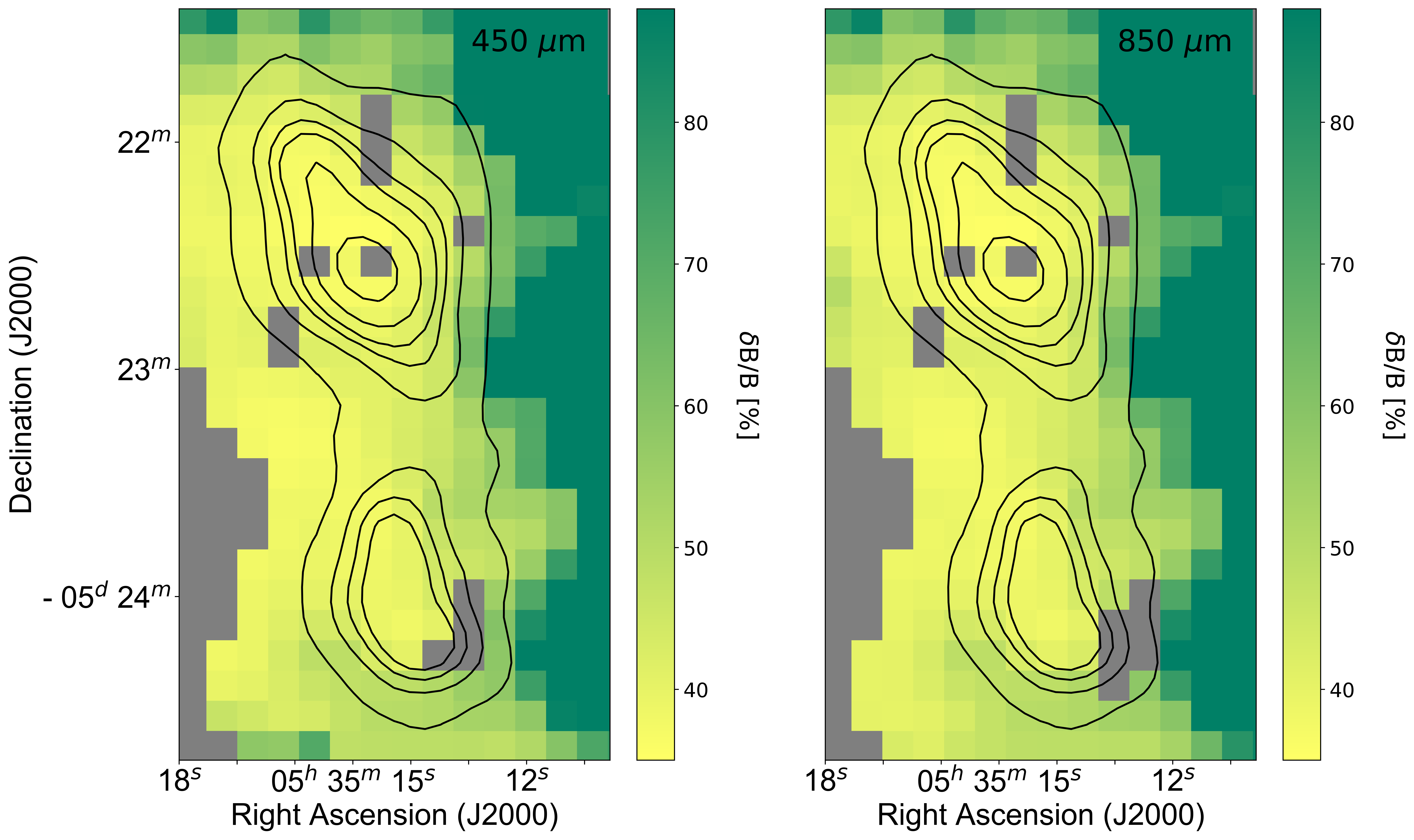}
\caption{The fractional uncertainties in magnetic field strengths in the OMC-1 region at 450 $\mu$m (left panel) and 850 $\mu$m (right panel). Contours in the map are the same as defined in Figure \ref{fig:f6}. \label{fig:f7}}
\end{figure*}

The uncertainty in volume density is estimated from the uncertainty in column density which is obtained using Equation (\ref{eq:bb}). In Equation (\ref{eq:bb}), a dominant systematic uncertainty is from the dust opacity $\kappa_{\nu_0}$. Its fractional uncertainty is about 50\% (e.g., \citealt{Roy2014}). The uncertainty in dust opacity index, $\delta \beta$, is in the range $\pm$0.3 (e.g., \citealt{Kwon2009}; \citealt{Schnee2010}; \citealt{Planck2011}; \citealt{Sadavoy2016}). The fractional calibration uncertainties at 450 $\mu$m and 850 $\mu$m are 10\% and 5\%, respectively \citep{Dempsey2013}. We assume a uniform depth over the OMC-1 region, when we calculate a volume density from the column density.  
We calculate the uncertainty in dust temperature as follows:

\begin{eqnarray} 
\frac{1}{T^2} \bigg(\frac{\frac{h\nu_{850}}{k_B}e^{h\nu_{850}/k_BT}}{e^{h\nu_{850}/k_BT} -1} - \frac{\frac{h\nu_{450}}{k_B}e^{h\nu_{450}/k_BT}}{e^{h\nu_{450}/k_BT} -1} \bigg )^2 \bigg(\frac{\delta T}{T}\bigg)^2 \nonumber \\
=\bigg (\frac{\delta I_{850}}{I_{850}} \bigg)^2+\bigg( \frac{\delta I_{450}}{I_{450}} \bigg ) ^2+ \bigg ( \delta \beta \ln \bigg(\frac{\nu_{850}}{\nu_{450}} \bigg) \bigg)^2,
\end{eqnarray}
where $\delta T$ is the uncertainty in dust temperature, $T$; $\delta  I_{850}$ and $\delta I_{450}$ are the uncertainties in flux densities at 450 $\mu$m and 850 $\mu$m, respectively. Other notations are the same as in Equation (\ref{eq:temp}). Consequently, the uncertainty in column density is dependent on the dust temperature. The obtained mean fractional uncertainty in column density is about 88\% inside the lowest contour level of the OMC-1 region (Figure \ref{fig:f7}), which is estimated using the error propagation Equations (\ref{eq:temp}) and (\ref{eq:bb}). If we simply assume that there is a 50\% fractional uncertainty in the depth determination, the mean fractional uncertainty in volume density estimation is 101.2\%.  It is about 13\% larger than the uncertainty of the column density estimate. If we include the uncertainty of the depth determination, the mean fractional uncertainty in magnetic field strengths is 50.6\%. Since the choice of 50\% uncertainty of the depth determination is arbitrary, we did not include the uncertainty of the depth estimate in the following sections of this paper. However, we note that the uncertainty of our determination of magnetic field strength is underestimated.

We consider the Gaussian fitting error to be appropriate for the uncertainty in velocity dispersion, $\delta \Delta V$. The mean fractional uncertainty in velocity dispersion inside the lowest contour level of the OMC-1 region map is about 1.2\%. We take the uncertainty in polarization angle dispersion from the measurement uncertainty in polarization angle. The polarization angle and its uncertainty is determined by Stokes $Q$, $U$ and their uncertainties. We calculate mean value of polarization angle uncertainties in the 5$\times$5 pixel box as the uncertainty of the measured polarization angle dispersion. 
The uncertainties in polarization angle range from 0.06 to 0.72 degree at both wavelengths inside the lowest contour level of the OMC-1 map. The mean fractional uncertainty in polarization angle dispersion is about 3.3\% in the same region.

We use Equation (\ref{eq:db_b}) to obtain the fractional uncertainties on magnetic field strengths in the OMC-1 region shown in Figure \ref{fig:f7}.  The dominant term in Equation (\ref{eq:db_b}) is the fractional uncertainty in volume density. The mean fractional uncertainties in the measurement of magnetic field strength within the lowest contour level in Figure \ref{fig:f7} are 44.7\% at 450 $\mu$m and 45.1\% at 850 $\mu$m.

\section{Discussion} \label{sec:disc}

\subsection{Magnetic field orientation and strength} \label{subsec:Bdisc}

The overall distributions of magnetic field strengths at 450 $\mu$m and 850 $\mu$m show good agreement and show two parts having relatively strong field strengths, the BN-KL clump and the region between two clumps. The BN-KL clump shows high magnetic field strength, which is due to high density there. 
Parts showing the lowest angle dispersion and highest magnetic field strength are located at a region between two clumps. In the region, magnetic field lines are highly ordered as shown in Figure \ref{fig:mag} as blue segments. \citet{Pattle2017} proposed that the approaching two BN-KL and S clumps could explain the hourglass morphology of the magnetic field lines between the clumps. In accordance with their picture, we think that the magnetic field lines are compressed by the two approaching clumps, so magnetic field strengths are strong in the region. Recently, \citet{Guerra2020} obtained maps of magnetic field strengths using four band data of SOFIA. The magnetic field strengths in their maps at 154 $\mu$m and 214 $\mu$m are also relatively strong in the two parts.

The mean magnetic field strength inside the lowest contour level at 450 $\mu$m and 850 $\mu$m is about 6 mG in the OMC-1 region, which is larger than those obtained by previous studies using the DCF method. For example, using their `structure function' method, \citet{Hildebrand2009} estimated a plane-of-sky magnetic field strength, 3.8 mG, in the OMC-1 region with the CSO Hertz polarimeter data. \citet{Chuss2019} estimated magnetic field strengths ranging from 0.9 to 1.01 mG in the OMC-1 region with SOFIA data using the dispersion function approach suggested by \citet{Houde2009}. Their result is similar to a magnetic field strength in the OMC-1 region, 0.76 mG, estimated by \citet{Houde2009}. 
Recently, \citet{Guerra2020} showed maps of magnetic field strengths obtained with four band data of SOFIA by applying the two-point structure function suggested by \citet{Houde2009} within a small circular region having a 9 pixel radius. The maximum value of their magnetic field estimations in the maps is 2 mG. All of these studies apply the DCF method in a larger region than our box size, which results in smaller magnetic field strengths than our measurements.

However, our mean field strength is consistent with the previous estimation by the BISTRO Survey data in the region within uncertainty \citep{Pattle2017}.  They estimated a plane-of-sky magnetic field strength in the OMC-1 region of 6.6 $\pm$ 4.7 mG with POL-2 data using their `unsharp masking' method. The box sizes used by them and us are similar. 

There are also other attempts to study a magnetic field in the OMC-1 region using single dish telescope and interferometers. There are earlier measurements using the BIMA array \citep{Rao1998} and the CSO \citep{Schleuning1998}. They showed that polarized directions in the region are well-ordered and the polarized emission comes from magnetically aligned dust grains. \citet{Vallee1999} measured the magnetic field at 8 positions in the OMC-1 region and found similar magnetic field orientations. \citet{Schleuning1998} estimated a magnetic field strength in the region,  about 1 mG. Some estimations of magnetic field strengths in the BN-KL were conducted using the Zeeman effect of the OH masers with the Very Large Array (VLA) \citep{Johnston1989} and the Multi Element Radio Linked Interferometer Network (MERLIN) \citep{Cohen2006}. The magnetic field strengths range from 1 to 16 mG, which are similar to our range of the strengths. \citet{Tang2010} estimated the magnetic field strength in the BN-KL using the SMA. It is larger than 3 mG based on a dynamical argument involving the explosive outflows in the region. The idea is that in the presence of ram pressure generated by outflows, clumps would disintegrate unless additional surface tension is generated by the presence of a magnetic field. This sets a lower limit to the necessary field strength. These measurements obtained by interferometers focus on the near young stellar objects in the BN-KL. Their size scale is about 200-2000 au, which is much smaller than our size scale of 0.1 pc, so these results show at denser and deeper region than our analyzed region.

\begin{figure*}[tbh!] 
\epsscale{1.0}
\plotone{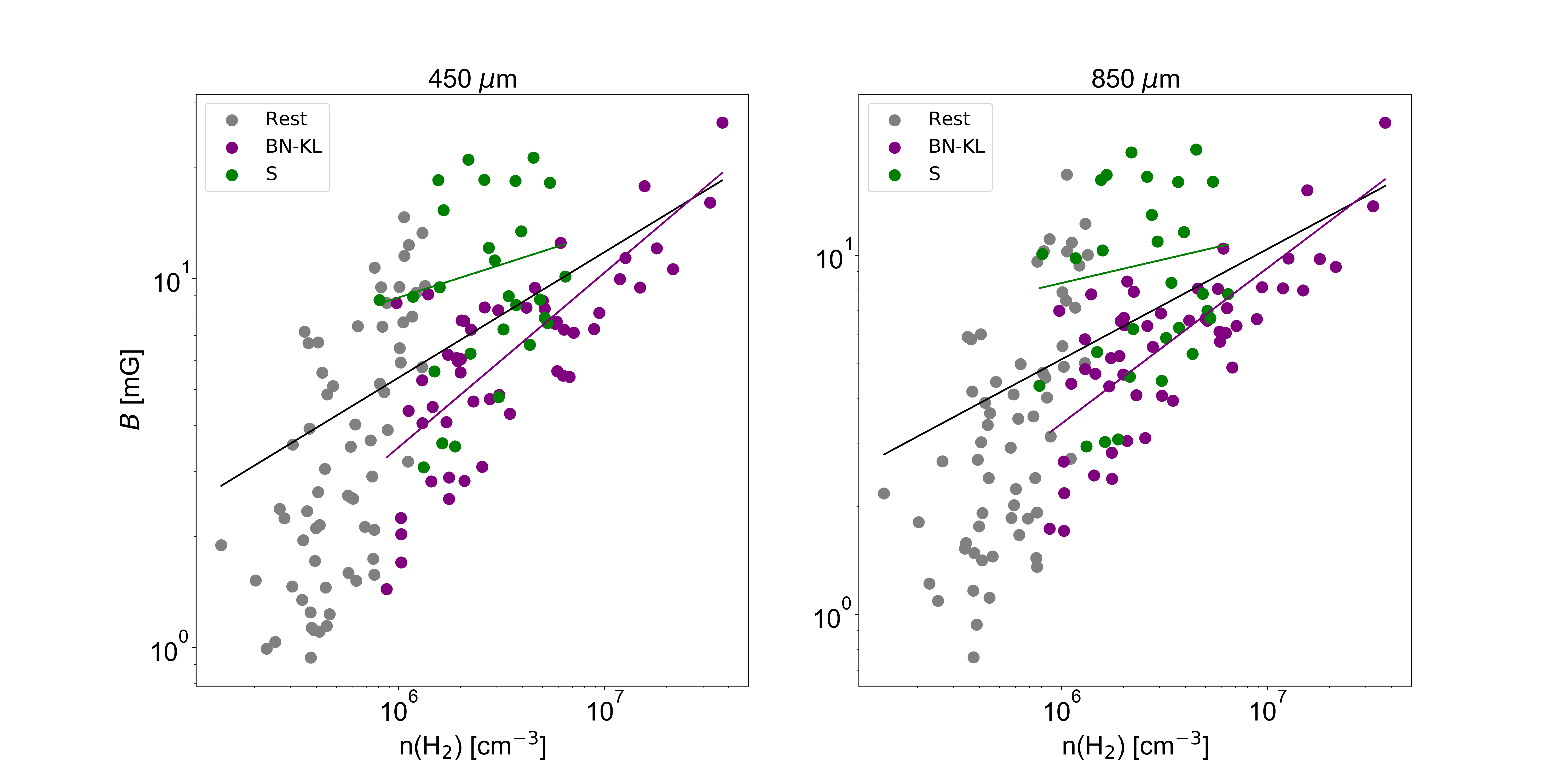}
\caption{Magnetic field strengths against volume densities at 450 $\mu$m (left panel) and 850 $\mu$m (right panel). 
Purple and green dots for the BN-KL and S clumps, respectively, which are inside regions defined with the second lowest contour level in Figure \ref{fig:mag}. Gray dots for the region defined by the lowest and second lowest contour levels in the figure.
Purple, green and black lines indicate the best-fit lines from the least-square regressions to the purple, green and all dots, respectively. The $\kappa$ values of all lines at both panels are shown in Table \ref{table:t_kappa}   \label{fig:nvsB}}

\end{figure*}

\subsection{$B \propto n(\text{H}_2)^\kappa$}

The magnetic field strengths in general increase with increasing density because magnetic field lines are dragged along materials in a molecular cloud by gravitational collapse. The magnetic field strength can be expressed with a power-law of volume density, $B \propto n(\text{H}_2)^\kappa$. \citet{Mastel1966} estimated $\kappa$ = 2/3 for a spherical collapse case with weak magnetic field strength. \citet{Crutcher2010} also estimated $\kappa \approx 0.65$ from magnetic field strengths obtained using Zeeman effect in interstellar clouds. However, a strong magnetic field can constrain collapse and make clouds flattened along the magnetic field. The cloud contraction across the magnetic field is driven by ambipolar diffusion. The $\kappa$ value of a cloud with ambipolar diffusion is known to be less than 0.5 \citep{Mouschovias1999}. 

We obtain the $\kappa$ values using volume densities and magnetic field strengths at 450 $\mu$m and 850 $\mu$m. Figure \ref{fig:nvsB} shows measured field strengths as a function of volume density at each wavelength. Purple and green dots for the BN-KL and S clumps, respectively, which are inside regions defined with the second lowest contour level in Figure \ref{fig:mag}. Gray dots for the region defined by the lowest and second lowest contour levels in the figure.
We perform the least-square fits to purple, green and the whole (purple, green and gray) dots, which correspond to the BN-KL, S clumps and the whole region, respectively.
 The best-fit $\kappa$ values are shown in Table \ref{table:t_kappa}. The $\kappa$ values in each region at both wavelengths are nearly the same with each other. Our results are consistent with the $\kappa$ value predicted from the ambipolar diffusion model. The $\kappa$ values in S clump are about twice smaller than those in other two regions.

\begin{deluxetable}{cccc}
\tablecolumns{4}
\tablewidth{0pt}
\tablecaption{$\kappa$ values 
\label{table:t_kappa}}
\tablehead{\colhead{Wavelengths ($\mu$m)}  &\colhead{$\kappa_{\text{BN-KL}}$} & \colhead{$\kappa_{\text{S}}$}& \colhead{$\kappa_{\text{all}}$} }
\startdata
450  & 0.47$\pm$0.043 & 0.18$\pm$0.19 & 0.34$\pm$0.035 \\
850  & 0.43$\pm$0.041 & 0.13$\pm$0.19 & 0.31$\pm$0.037
\enddata
\vspace{-0.8cm}
\end{deluxetable}

We, however, note that there is a couple of caveats in the above interpretation. First, we use estimated magnetic field strength in the plane of the sky rather than three dimensional magnetic field strength. Second, we roughly estimate volume densities from column densities.

\subsection{Mass-to-flux ratio}\label{subsec:mtob}

We measure the observed mass-to-flux ratio, $(M/\Phi)_{\rm obs}$, which is often used to determine whether or not magnetic fields can support a molecular cloud against gravitational collapse (\citealt{Mouschovias1976}; \citealt{Crutcher2004}). The observed mass-to-flux ratio $\lambda$ in units of the critical ratio  of a magnetized disk of which mass is marginally supported by a magnetic field, $(M/\Phi)_{\rm crit}=1/2\pi G^{1/2}$ \citep{Nakano1978}, is as follows \citep{Crutcher2004};
\begin{equation}
\lambda \equiv \frac{(M/\Phi)_{\text{obs}}}{(M/\Phi)_{\text{crit}}} = \frac{\mu m_{\text{H}} N(\text{H}_2)/B}{1/2\pi G^{1/2}} = 7.6 \times10^{-21} \frac{N(\text{H}_2)}{B},\label{eq:mass_flux_ratio}                                            
\end{equation}
where $B$ is the strength of three-dimensional magnetic field in $\mu$G and $N$(H$_2$) is in cm$^{-2}$. A value of $\lambda < $ 1 means that the molecular cloud is magnetically subcritical, so the magnetic field could prevent gravitational collapse of the molecular cloud. $\lambda >$ 1 implies the cloud is magnetically supercritical, so magnetic field cannot resist the gravitational collapse.

\begin{figure*}[thb!]
\epsscale{1.1}
\plotone{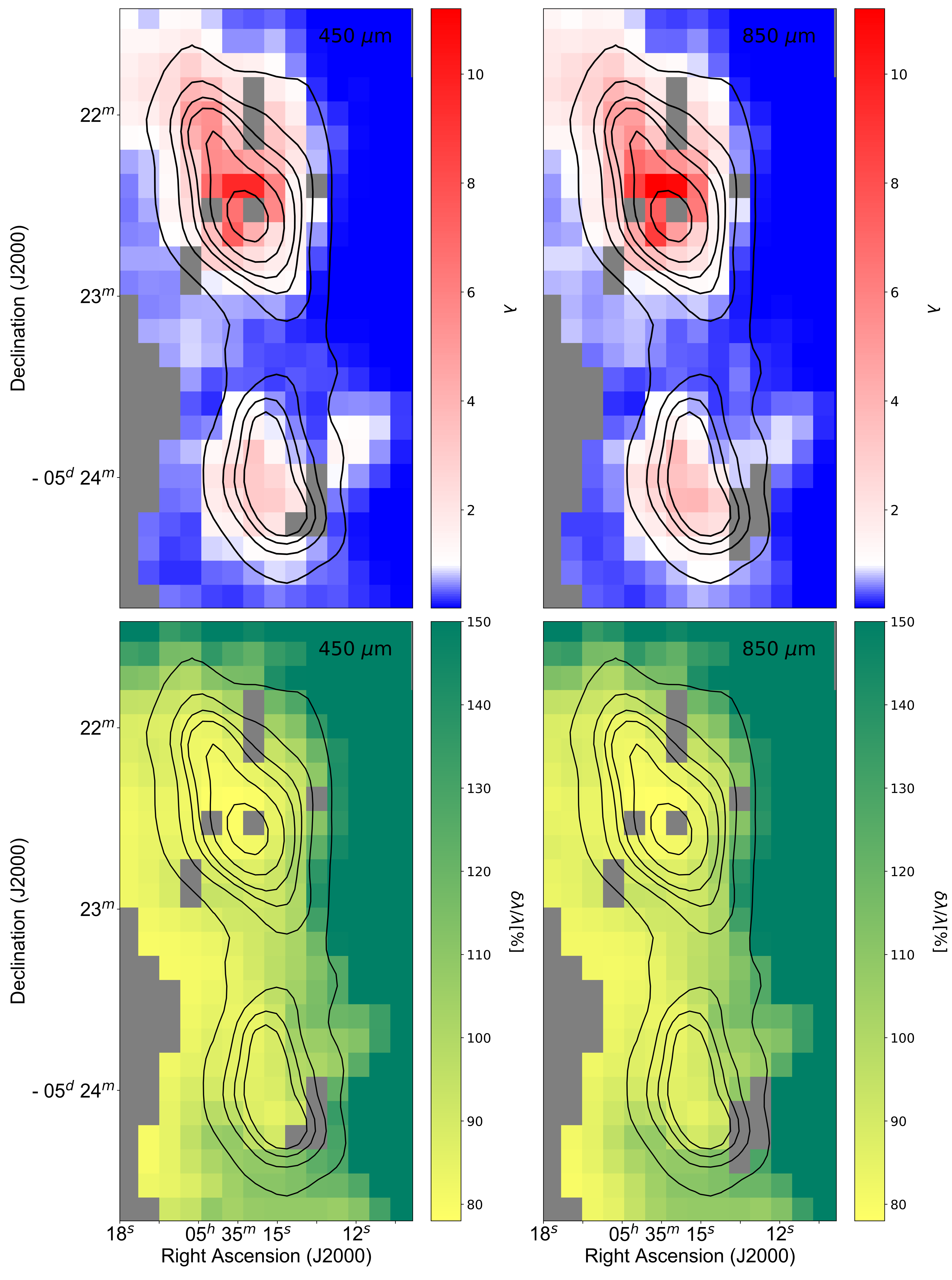}
\caption{Maps of the mass-to-flux ratios in units of the critical ratio (upper panels) and their uncertainties (lower panels) in the OMC-1 region at 450 $\mu$m (left panels) and 850 $\mu$m (right panels). Black contours are as defined in Figure \ref{fig:f6}. Note that the polarization angle dispersion for the DCF method is obtained in a 40$''$ box and the box moves by 8$''$ over the region. \label{fig:mtob}}
\end{figure*}

To estimate the mass-to-flux ratio, we should know the strength of the three-dimensional magnetic field. However, we can only estimate the distribution of the plane-of-sky magnetic field strengths in the OMC-1 region using the DCF method. The magnetic field strengths along the line of sight have been estimated by the measurement of the Zeeman effect. There are a number of Zeeman measurements of CN, OH and HI spectral lines towards the OMC-1 region, most of which fall in the vicinity of a high-extinction region. The magnetic fields inferred from these different studies have wildly different values and, often, large error bars; for example, 360 $\pm$ 80 $\mu$G (\citealt{Falgarone2008}; \citealt{Crutcher1999}; \citealt{Crutcher2010}), 79 $\pm$ 99 $\mu$G \citep{Crutcher1996}, 40 $\pm$ 240 $\mu$G \citep{Crutcher2010}, and 80 $\pm$ 100 $\mu$G \citep{Crutcher2010}. These studies suggest that the the line-of-sight magnetic field in this region (including error bars) might have a maximum strength, 440 $\mu$G,  which is smaller than 25\% of the minimum plane-of-sky magnetic field strength in the OMC-1 region. The magnetic field strength along the line-of-sight is weaker than that in the plane of the sky, which is consistent with the bow-shaped magnetic field morphology suggested by \citet{Tahani2019}. \citet{Tahani2018} estimated magnetic field strengths in the Orion A cloud using a new technique based on rotation measure. Among the data in their Orion A map, the locations of two measurements of the line-of-sight magnetic field strengths, 23 $\pm$ 38 $\mu$G and 15 $\pm$ 36 $\mu$G, are close to our analyzed region. However, we note that these values in \citet{Tahani2018} are averaged along the line of sight within the Orion A molecular cloud. Recently, \citet{Guerra2020} obtained the maps of the line-of-sight magnetic field strengths from the maps of angle dispersion. The field strengths are up to 2 mG nearby the BN-KL clump which is smaller than 20\% of our estimated magnetic field strengths at the same location. The magnetic field strengths in the plane of the sky are far larger than the field strengths along the line-of-sight. Therefore, we approximate a magnetic field strength in the sky plane as a three-dimensional magnetic field strength.  

Figure \ref{fig:mtob} shows the distributions of the mass-to-flux ratio in units of the critical ratio and their uncertainties across the OMC-1 region at 450 $\mu$m and 850 $\mu$m. The two clumps appear magnetically supercritical, while the parts between the two clumps are magnetically subcritical. Since the plane-of-sky magnetic field strength corresponds to a lower limit of the three-dimensional magnetic field strength, the subcritical region would not change if the line-of-sight component is added in. The effect of possibly underestimating the total field strength by not taking into account the line-of-sight component is 
that the supercritical area in the figure might shrink, i.e., become more centered only on the highest emission peaks. 
The mass to flux ratios vary from 0.2 to 9.5 at 450 $\mu$m and from 0.3 to 11.2 at 850 $\mu$m. Their mean and median values of mass-to-flux ratios inside the lowest contour level shown in Figure \ref{fig:mtob} are 1.9 and 1.4 at 450 $\mu$m and 2.1 and 1.5 at 850 $\mu$m, respectively. The uncertainty of the mass to flux ratio is estimated using the uncertainties of the column density and magnetic field strength. The mean fractional uncertainties of the ratio inside the lowest contour are about 93\% at both wavelengths. When one interprets the maps of magnetic field strength (Figure \ref{fig:mag}) and mass-to-flux ratio (Figure \ref{fig:mtob}), the constant depth assumption should be taken into account. Since, observationally, there is no good way to get physical extents of the OMC-1 region along different sight lines, we simply assume a constant depth. The depth estimation is done by comparing the column and volume densities at the column-density peak position of the BN-KL clump. As we explained in section \ref{subsec:col}, the implication of the constant depth assumption is that an effective volume density of polarized dust emission layers at a specific sight line is proportional to the column density at the same sight line.

Two possible mechanisms, ambipoar diffusion and magnetic reconnection, may explain the increasing trend of the mass-to-flux ratio from low to high densities seen in our results. Of the two, the ambipoar diffusion is the most accepted mechanism.  The ambipoar diffusion enables neutral particles to move across magnetic field lines.  The diffusion process is accelerated when the degree of ionization in a high density environment becomes small. The magnetic reconnection in a static two-dimensional geometry is a slow process and the scale of annihilation of magnetic field is small \citep{Shu1991}. However, magnetic reconnection in a turbulent three-dimensional environment is fast \citep{Lazarian1999}, which plays a certain role in the redistribution of mass-to-flux ratio in star formation context. We think that more detailed studies are needed especially on the length and time scales of the magnetic reconnection in the star formation context.

The mean values of the mass-to-flux ratios estimated by us are comparable to $\lambda \sim 2.2$ estimated by \citet{Crutcher1999} and larger than $\lambda \sim 0.41$ measured by \citet{Pattle2017}.  
 \citet{Pattle2017} estimated the mass-to-flux ratio from the median column density in the OMC-1 region. The column density is about five times smaller than our mean column density, so their mass-to-flux ratio is smaller than our mean values.

 The advantage of our method is to enable us to derive the distribution of mass-to-flux ratios across the OMC-1 region. The distribution clearly shows that the two clumps in the OMC-1 region are gravitationally unstable and have undergone active star formation. But the outer parts are supported by the magnetic field. This result suggests that, on the large scale, the OMC-1 region is a magnetically subcritical environment. The overall evolution of the region might be magnetically mediated. \citet{Pattle2017} hypothesized that the magnetic field in the region between the two clumps has been compressed by large-scale motions of material. Their picture could explain the interclump region is in a magnetically subcritical state. Recently, \citet{Guerra2020} obtained the maps of the mass-to-flux ratio of the OMC-1 region from the SOFIA data. They showed that the interclump region in all the maps is magnetically subcritical. They however showed inconclusive results that the mass-to-flux ratio around the BN-KL clump is magnetically subcritical at long wavelengths (154 $\mu$m and 214 $\mu$m), but it is magnetically supercritical at short wavelengths (53 $\mu$m and 89 $\mu$m).

\begin{figure*}[thb!]
\epsscale{1.0}
\plotone{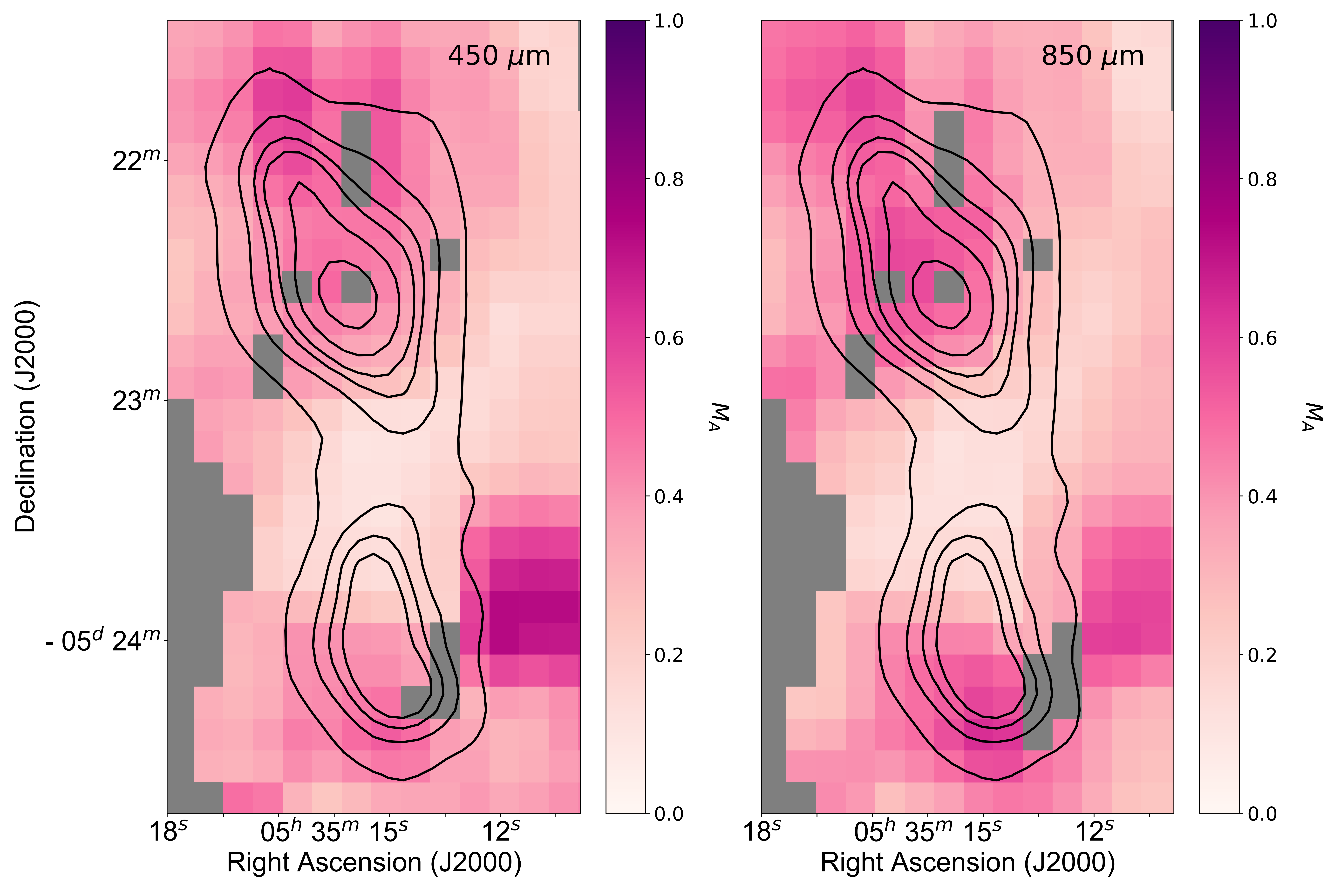}
\caption{Maps of the Alfv$\acute{e}$n Mach number in the OMC-1 region at 450 $\mu$m (left panel) and 850 $\mu$m (right panel). Contours are the same as defined in Figure \ref{fig:f6}. Note that the polarization angle dispersion for the DCF method is obtained in a 40$''$ box and the box moves by 8$''$ over the region. \label{fig:mach}}
\end{figure*}

\subsection{Alfv$\acute{e}$n Mach number}

Another parameter to study the role of magnetic field strength in star-forming region is the Alfv$\acute{e}$n Mach number ($M_A$).
It is the ratio of the turbulent velocity to the Alfv$\acute{e}$n speed. It has been used to assess the relative importance of the turbulence with respect to the magnetic field in molecular clouds \citep{Crutcher1999}. It is expressed as
$M_A=\sqrt{3}\sigma_v/v_A$, where $v_A = B/\sqrt{4\pi\rho}$ is the Alfv$\acute{e}$n speed. By assuming the plane-of-sky magnetic field strength is equal to the three-dimensional field strength, $B \equiv B_{\text{pos}}$, the Alfv$\acute{e}$n Mach number is simply written as $M_A=\sqrt{3}\sigma_\theta/Q$. The Alfv$\acute{e}$nic Mach number is proportional to the polarization angle dispersion. The sub-alfv$\acute{e}$nic condition, $M_A <$ 1.0,  means magnetic pressure exceeds turbulent pressure, and the super-alfv$\acute{e}$nic condition, $M_A  > 1.0$,  means turbulent pressure is more dominant than magnetic pressure in a molecular cloud. 
 
Figure \ref{fig:mach} shows the maps of the Alfv$\acute{e}$n mach number in the OMC-1 region at 450 $\mu$m and 850 $\mu$m. Inside of the lowest contour level, Alfv$\acute{e}$n mach numbers are smaller than 0.6 at both wavelengths and their mean values are $\sim 0.4$. So the OMC-1 region is subalfv$\acute{e}$nic, and magnetic pressure of the region is greater than the turbulent pressure. Magnetic field is relatively dominant compared to turbulence and can regulate star-forming processes in the region. 
The Alfv$\acute{e}$n mach number is dependent upon the angle dispersion, so the fractional uncertainty of the Alfv$\acute{e}$n mach number is the same as that of the angle dispersion. The uncertainty of measuring the  Alfv$\acute{e}$n mach number is just few percents, which is, in fact, the uncertainty of the polarization angle dispersion.  
 
\section{Summary} \label{sec:summ}

We propose a new application of the DCF method to estimate the distribution of magnetic field strengths in a molecular cloud or core. We apply it to a well-known star-forming region, the OMC-1 region. We use observations of polarized dust emission and C$^{18}$O spectral lines obtained from SCUBA-2/POL-2 and  HARP on the JCMT. 
Previous studies have measured a mean magnetic field strength over the whole of, or quite a large area of, a molecular cloud or core using the DCF method. Instead, we estimate the distribution of magnetic field strengths of the OMC-1 region at various locations using the following procedure:  

Firstly, we obtained a mean direction of polarization segments within a small box of 25 pixels, 40$\arcsec\times$40$\arcsec$. By moving the box over the whole OMC-1 region, we evaluate the distribution of mean angles. Second, we calculate the difference in angle between the observed and the estimated mean angles in each pixel. Then, we calculate the polarization angle dispersion in each box by taking the root mean squared of the angle differences. Lastly, substituting these values of volume density, velocity dispersion and angle dispersion in each box into the DCF formula, we obtain the distribution of magnetic field strengths.

The estimated magnetic field strengths in the plane of the sky range from 0.8 to 26.4 mG at 450$\mu$m and 850 $\mu$m and their mean values are 6.6 $\pm$ 3 mG at 450 $\mu$m and 6.2 $\pm$ 2.8 mG at 850 $\mu$m. These uncertainites in the means are based solely on the range of inferred values over the region and not on the uncertainties in deriving those values. The maps of magnetic field strengths at both wavelengths are quite consistent within uncertainties. The strongest magnetic field strength is in the inter region between the BN-KL and S clumps, and the magnetic field in that region is also highly ordered. 

The magnetic field strengths increase with column densities following a power-law relation in the OMC-1 region. The index of the relation is less than 0.5 in the region. It is consistent with the index expected in a cloud contracted by ambipolar diffusion. We additionally estimate the distribution of the mass-to-magnetic flux ratio in the OMC-1 region, which tells us whether or not the magnetic field in the region could prevent gravitational collapse. We assume the magnetic field in the plane of the sky is the dominant component of the three-dimensional magnetic field. The mass-to-flux ratios indicate that the central parts of the two clumps in the OMC-1 region are magnetically supercritical and the outer parts are magnetically subcritical. Based on this analysis, we expect that the central parts of the two clumps in the OMC-1 region are undergoing gravitational collapse, while the rest of the region is supported by the magnetic field. We also showed the distribution of Alfv$\acute{e}$n Mach number in the OMC-1 region. The mean Alfv$\acute{e}$n Mach number over the OMC-1 region is about 0.4, which means the magnetic pressure is stronger than the turbulent pressure but in the two clumps gravity is the dominant force in the region. 

\acknowledgments
W.K. was supported by the New Faculty Startup Fund from Seoul National University and by the Basic Science Research Program through the National Research Foundation of Korea (NRF-2016R1C1B2013642). C.L.H.H. acknowledges the support of the NAOJ Fellowship and JSPS KAKENHI grants 18K13586 and 20K14527. D.J. is supported by the National Research Council of Canada and by a Natural Sciences and Engineering Research Council of Canada (NSERC) Discovery Grant. R.S.F. is supported by JSPS KAKENHI grant 19H01938. C.W.L. is supported by Basic Science Research Program through the National Research Foundation of Korea (NRF) funded by the Ministry of Education, Science and Technology (NRF-2019R1A2C1010851). T.L. is supported by international partnership program of Chinese academy of sciences grant No.114231KYSB20200009. M.T. is supported by JSPS KAKENHI grant Nos.18H05442,15H02063, and 22000005. J.K. is supported JSPS KAKENHI grant No.19K14775. G.P. is supported by Basic Science Research Program through the National Research Foundation of Korea (NRF) funded by the Ministry of Education (NRF-2020R1A6A3A01100208). D.L. is supported from NSFC No 11911530226 and 11725313. The James Clerk Maxwell Telescope is operated by the East Asian Observatory on behalf of The National Astronomical Observatory of Japan; Academia Sinica Institute of Astronomy and Astrophysics; the Korea Astronomy and Space Science Institute; Center for Astronomical Mega-Science (as well as the National Key R\&D Program of China with No. 2017YFA0402700). Additional funding support is provided by the Science and Technology Facilities Council of the United Kingdom and participating universities in the United Kingdom and Canada.

\software{KAPPA \citep{Currie2008}, Starlink \citep{Jenness2013}, GILDAS/CLASS (\citealt{Pety2005}; \citealt{gildasteam2013}) }

\clearpage

\appendix

\section{Maps of the differences of polarization angles and angle dispersions}\label{sec:app0}

\begin{figure}[htb!]
\epsscale{0.7}
\plotone{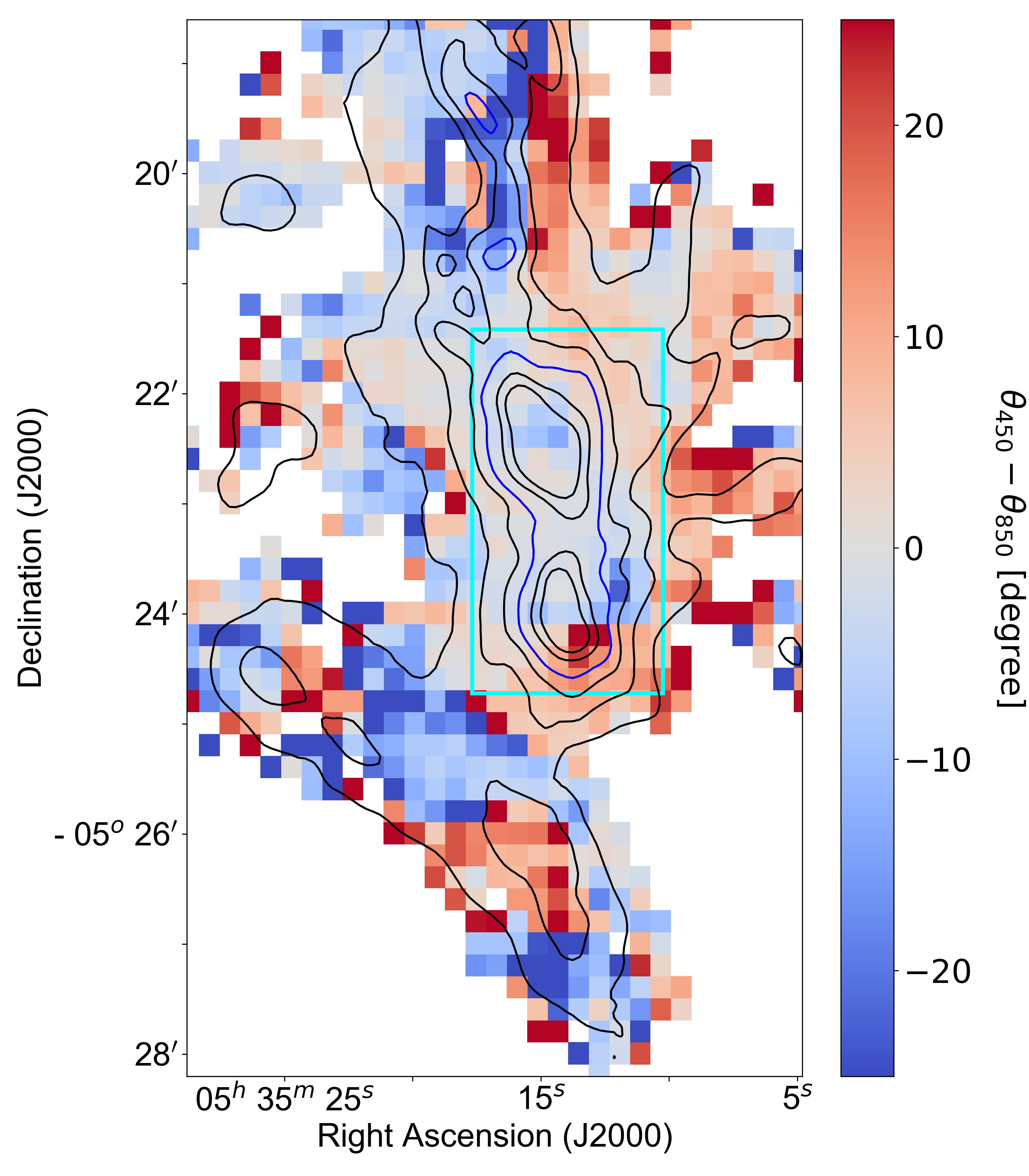}
\caption{The map of the differences of polarization angles at 450 $\mu$m and 850 $\mu$m. The cyan box is the same with the blue box shown in Figure \ref{fig:pol}. Contour lines trace equal intensities at 850 $\mu$m, whose values are 0.6, 1.8, 3.6, 6, 12 and 20 Jy beam$^{-1}$. Our analyzed region is inside the blue contour level, whose intensity is 6 Jy beam$^{-1}$. \label{fig:angdiff}}
\end{figure}

Most of polarization angles at 450 $\mu$m and 850 $\mu$m show quite good agreement within $\pm$ 25 degrees as shown in Figure \ref{fig:polang}. The map of angle differences at two wavelengths is shown in Figure \ref{fig:angdiff}. The mean values of all angle differences over the whole region and inside the cyan box are 0.5 and -0.6 degrees, and their standard deviations are 20.5 and 7.0 degrees, respectively. The mean measurement uncertainties of all polarization angles at 450 $\mu$m and 850 $\mu$m are 6.5 and 2.5 degrees. Most of angle differences inside the blue contour level are within 2.5 degrees which is smaller than the uncertainties. At few pixels at south-western part inside the contour level, angle differences are larger than 30 degrees or smaller than -20 degrees. These pixels are excluded in our analysis (see Appendix \ref{sec:appendixB}).

\begin{figure}[htb!]
\epsscale{0.5}
\plotone{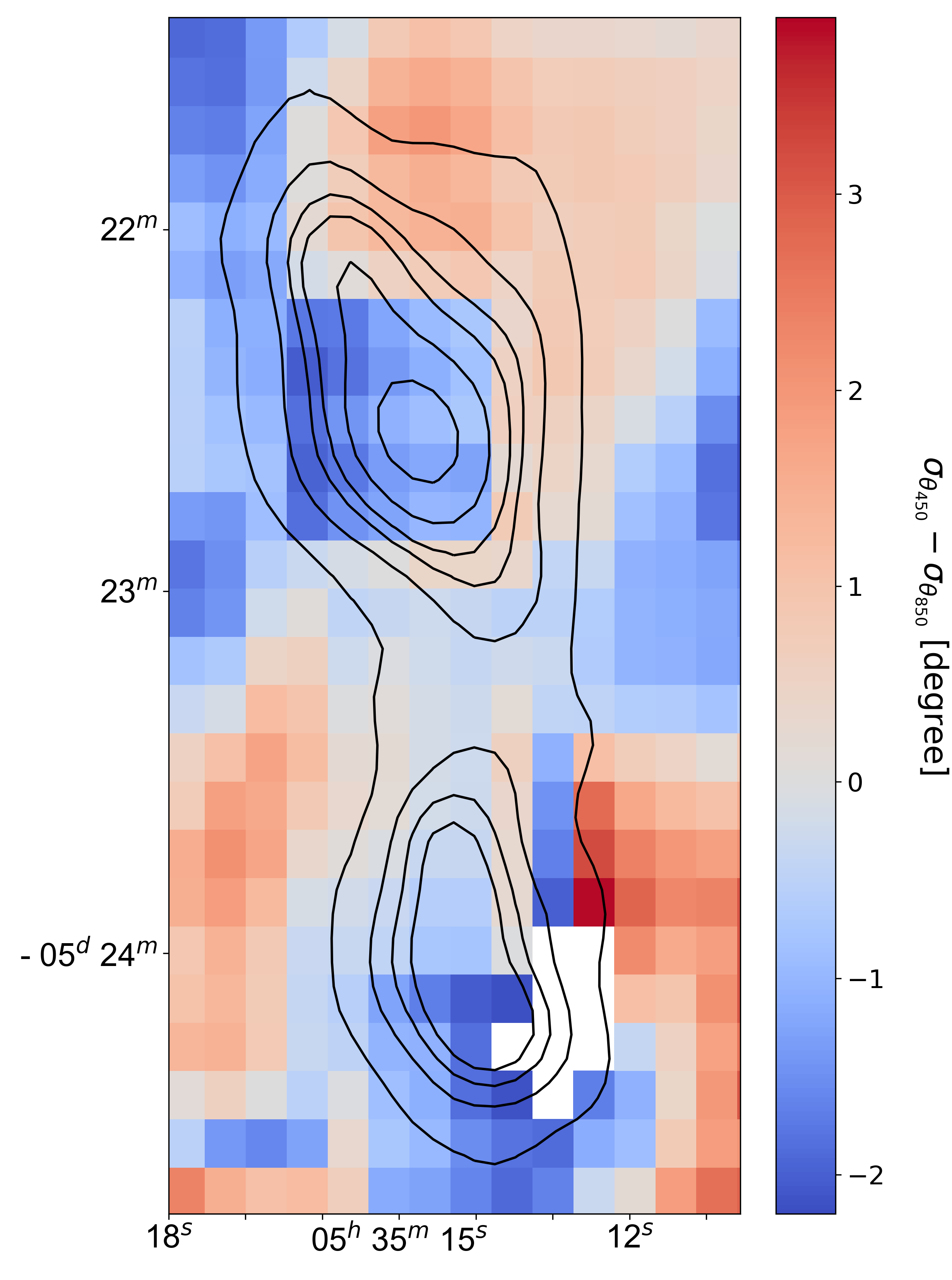}
\caption{The map of the differences of angle dispersions at 450 $\mu$m and 850 $\mu$m. Contours are the same as defined in Figure \ref{fig:f6} \label{fig:angdispdiff}}
\end{figure}

Figure \ref{fig:angdispdiff} shows the map of differences of angle dispersions at 450 $\mu$m ($\sigma_{\theta_{450}}$) and 850 $\mu$m ($\sigma_{\theta_{850}}$). The maps of angle dispersions at two wavelengths are shown in Figure \ref{fig:f6}. The mean and standard deviation values of the differences are -0.3 degrees and 1.1 degrees. Most of angle dispersions at two wavelengths are similar within $\pm$ 2 degrees which is smaller than the mean measurement uncertainties of polarization segments at 450 $\mu$m and 850 $\mu$m.

\section{Angle dispersion using a small moving box}\label{sec:appendixA}

\begin{figure*}[bht!]
\epsscale{0.8}
\plotone{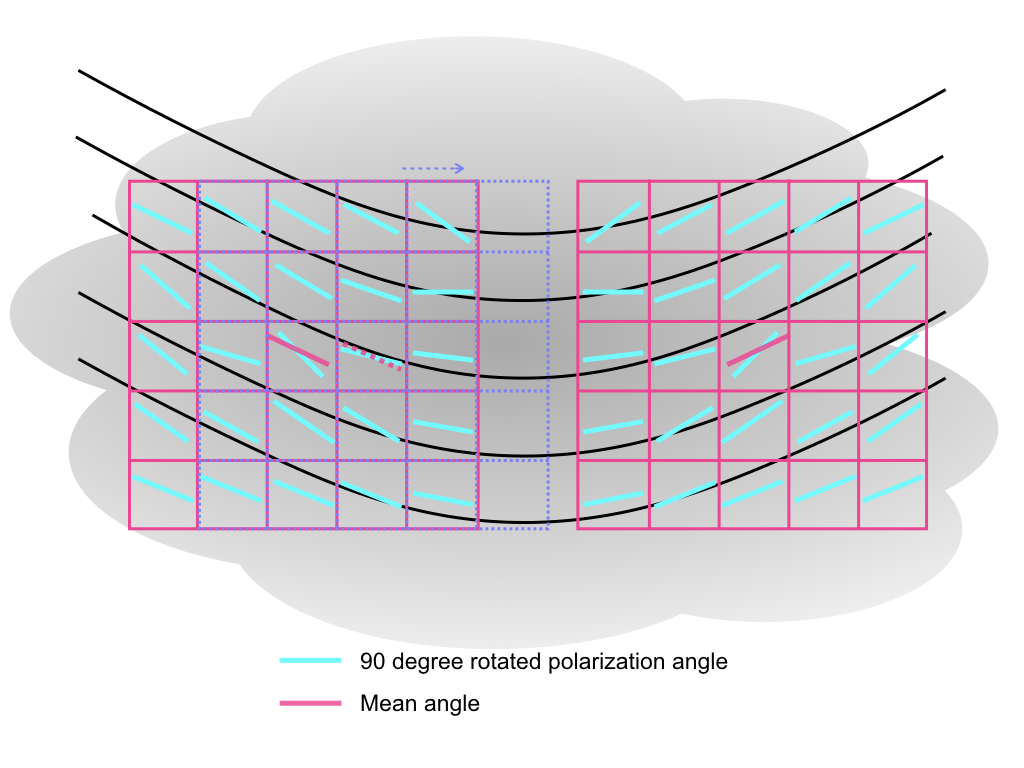}
\caption{A cartoon to explain the estimation of mean magnetic field orientations in a molecular cloud. The black lines represent magnetic field lines. An average field orientation is calculated inside a pink box whose size is 5$\times$5 pixels. Individual pixels have a polarization segment shown in cyan. The mean direction of 25 segments is drawn at the central pixel of the box in pink. After getting a mean direction in the box, we move the box by a pixel, represented by dashed lines. \label{fig:cartoon}}
\end{figure*}

We explain the detailed processes to estimate the angle dispersion by which we manipulate our polarization data to obtain the mean field distribution in OMC-1 using a box containing 25 pixels. Figure \ref{fig:cartoon} shows a cartoon of our method. The moving box is represented by a 5$\times$5 pixel box (in pink) and each pixel contains a model magnetic field segment (e.g., a polarization segment rotated by 90 degrees). We calculate a mean polarization angle, $\overline{\theta} = 0.5\arctan(\overline{U}/\overline{Q})$ in the box, where $\overline{U}$ and $\overline{Q}$ are the mean values averaged over the box. The obtained mean polarization angle is assigned to the central pixel of the box. We repeat the calculation of the mean polarization angle by moving the box over the whole OMC-1 region by one pixel at a time.
After obtaining the map of mean polarization angles in the OMC-1 region, we calculate the angle difference between each original segment and the mean polarization angle at the position, i.e. $\delta \theta _{i,j} = \theta _{i,j} - \overline{\theta}_{i,j}$, at pixel ($i$, $j$). Due to the 180-degree ambiguity of the polarization angle, we take the angle between the original and the mean polarization segments to be less than 90 degree by assuming that the magnetic field orientations do not dramatically change in the small box, 40$\arcsec\times40 \arcsec$. We exclude polarization segments if their directions make significant changes in the box based on radii of curvature of the segments (see Appendix \ref{sec:appendixB}). Since we know an angle difference at every pixel of the OMC-1 region, we calculate an angle dispersion in the box as the root mean squared of the angle differences in 25 pixels, $\sqrt{\sum{ \delta \theta _{i,j} ^2}/25}$. The angle dispersion obtained is assigned to the central pixel of the box. By moving the box one pixel at a time and calculating the angle dispersion in the box over the whole OMC-1 region, we derive an angle dispersion map of the region. 

\section{Curvature of magnetic field lines}\label{sec:appendixB}

\begin{figure*}[htb!]
\epsscale{1.0}
\plottwo{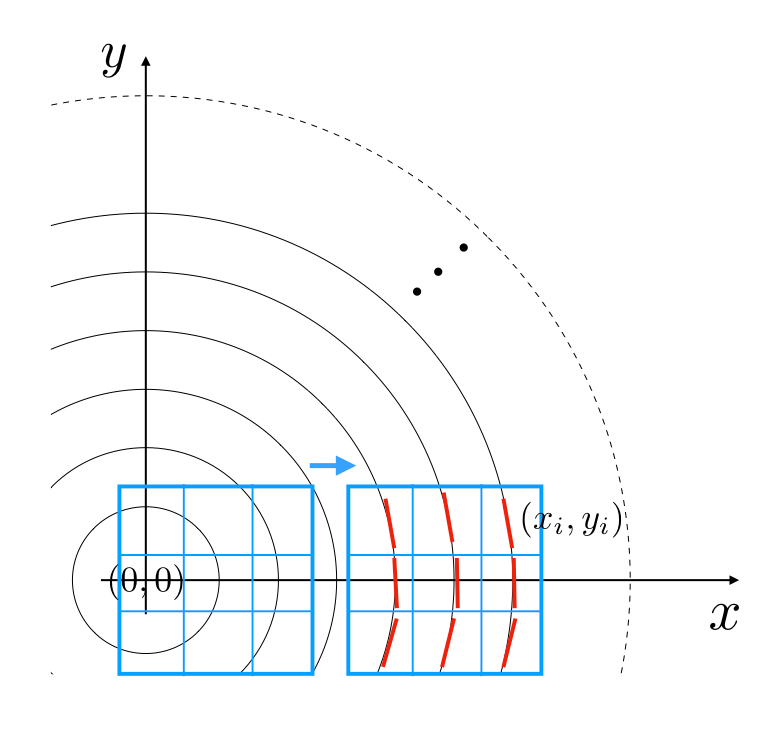}{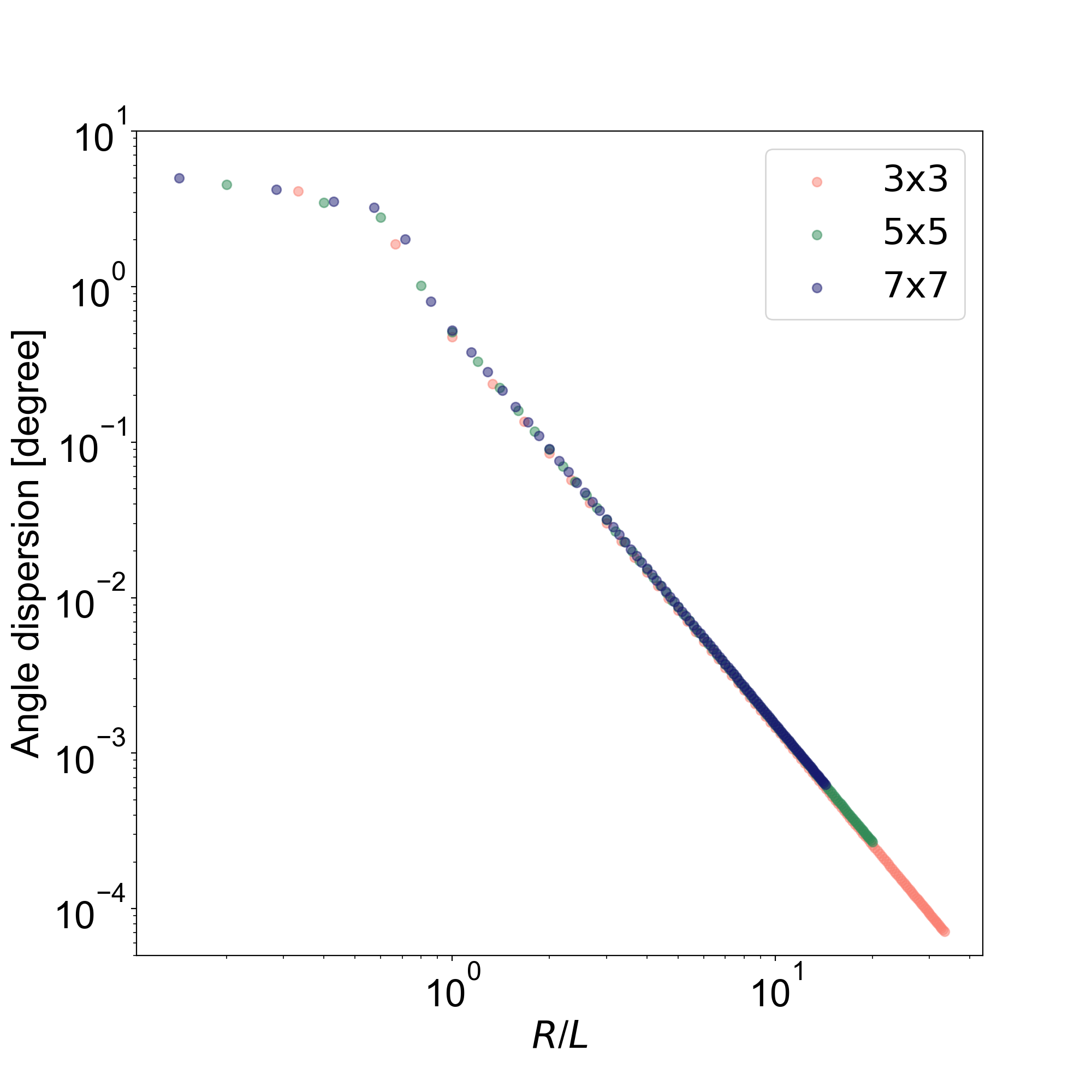}
\caption{Concentric circular field lines and averaging boxes (left panel). The black lines represent concentric circles. The origin of the circles is (0,0). 3$\times$3 pixel boxes are represented by thick blue lines, in which angle dispersions are calculated. The box moves along $x$-axis. Red segments are tangent lines to the circles. The angle dispersions as a function of the ratio of the radius of curvature to the box size using three different boxes (right panel). Red, green and blue dots are obtained by 3$\times$3, 5$\times$5 and 7$\times$7 pixel boxes, respectively.\label{fig:model}}
\end{figure*}

It is necessary for us to justify which box size is appropriate to estimate the angle dispersion in the OMC-1 region, so we examine the relation between box size and radius of curvature of magnetic field lines described by polarization segments. We calculate the change of angle dispersion by increasing the radius of curvature for three different box sizes in a very idealized set of concentric circular magnetic field lines. The concentric circles, with a common origin at (0,0) in the ($x,y$) plane, are shown as black lines in the left panel of Figure \ref{fig:model}. A radius of curvature at an ($x_i,y_i$) position is the same as the radius of a circle which passes through the ($x_i,y_i$) point. A tangent line of the circle at that position is expressed as a red segment in the figure. The slope of the line is expressed as an angle. The blue lines show a 3$\times$3 pixel box. We calculate a mean angle of nine slopes in the box and assign the mean angle to the center pixel of the box.  After determining mean angles in the nine pixels by moving the box vertically, horizontally and diagonally by one pixel, we estimate the angle difference between the mean angle and the original angle of a slope at each pixel. We then take the root mean squared of the nine angle differences as the angle dispersion in the box and assign the angle dispersion to the central pixel of the box. We calculate angle dispersions by moving the center pixel of a box along the $x$-axis from one-pixel distance from the origin to 100-pixel distance. These processes are also taken 
in the observational analysis to estimate the distribution of angle dispersion (see section \ref{subsec:angdisp}). We repeat the previous processes over the same concentric circular field lines for 5$\times$5 and 7$\times$7 pixel boxes to estimate angle dispersions as a function of the ratio of the radius of curvature to the box size ($R/L$).
 
 The right panel of Figure \ref{fig:model} shows estimated angle dispersions as a function of $R/L$ using three different boxes in units of pixels. The angle dispersion decreases as a function of $R/L$. When the ratio is infinite such as when the box size is very much smaller than the radius of curvature ($R$/$L$ $\gg$ 1), the angle dispersion should approach zero. Indeed, when the box size is smaller than the radius of curvature ($L < R$), the angle dispersion is quite well estimated. As the box size approaches the radius of curvature (e.g., $L \sim R$), the angle dispersion increases rapidly.  For example, at $L = R/2$, the angle dispersion is $<$ 0.1 degrees, whereas at $L = R$, the angle dispersion is a factor of five higher at $\sim$ 0.5 degrees.
 
\begin{figure*}[thb!]
\epsscale{1.0}
\plotone{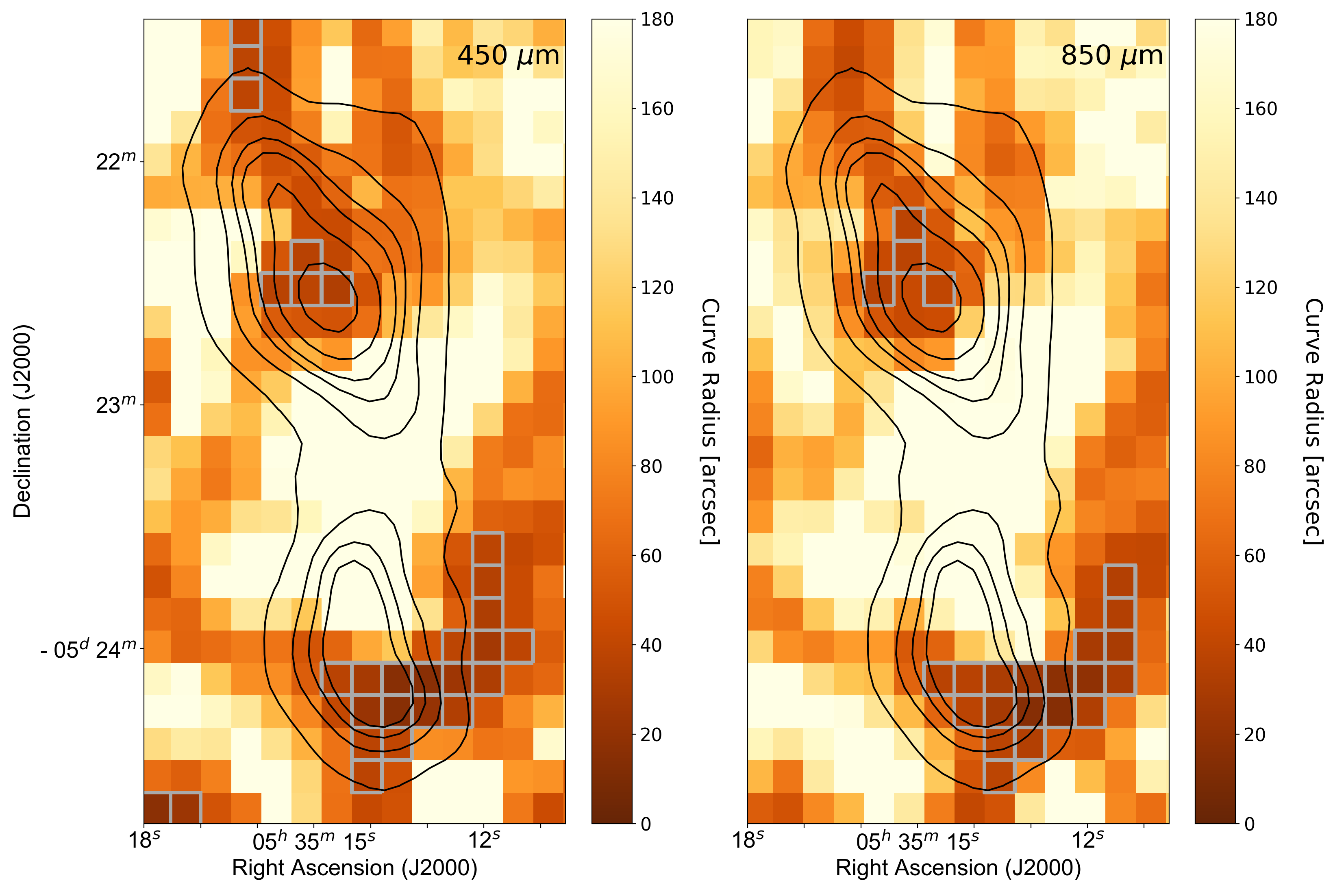}
\caption{The radius of curvature maps at 450$\mu$m (left panel) and 850$\mu$m (right panel) measured in units of arcseconds using polarization segments at both wavelengths in the OMC-1 region. The radius in each pixel is the mean value of four radii of curvature obtained from adjacent segments. The pixels outlined in gray, whose curvature radii are smaller than 40$''$, are excluded in the calculation of angle dispersion. \label{fig:f9}}
\end{figure*}

In our polarization map of the OMC-1 region, we calculate a radius of curvature at every measured point. A radius of curvature can be estimated by drawing a circle that goes through the two adjacent pixel points at which the two segments become tangent lines to the circle. The curvature radius will be increased when two polarization segments have similar directions. \citet{Koch2012} estimated the curvature of a magnetic field line that goes through two neighboring polarization segments using the following formula:

\begin{equation}
C \equiv \frac{1}{R} = \frac{2}{d}\cos\bigg (\frac{1}{2}\ [ \pi - \Delta PA ] \bigg),
\end{equation}
where $C$ is the curvature, $R$ is the radius of curvature, $d$ is the distance between two neighboring segments and $\Delta PA$ is the difference of polarization angles at two adjacent pixels. Four pairs of polarization segments are available at a pixel with its left, right, up and down pixels. The mean value of the four radii of curvature obtained from the pairs is assigned to the pixel.

 By applying this measurement of the radius of curvature at every point, we obtain radius of curvature maps of the OMC-1 region at 450 $\mu$m and 850 $\mu$m in units of arcseconds (Figure \ref{fig:f9}). The pixels outlined in gray have radii of curvature $<$ 40$''$, which is the same as a side of a 5$\times$5 pixel box. All radii of curvature except those in gray pixels inside the lowest contour level are larger than the box size. A 3$\times$3 pixel box also can be used to estimate angle dispersion in the OMC-1 region, but the box contains only nine pixels. It is not enough to estimate mean angle and angle dispersion.  Also, a side of the box is 24$''$ which is only about 1.6 times larger than the JCMT beam size at 850 $\mu$m. It only contains less than three beams. Therefore, we choose the 5$\times$5 pixel box to measure angle dispersion and magnetic field strength distribution in the OMC-1 region.

\end{document}